\documentclass[12pt]{article}
\usepackage{latexsym}
\usepackage{amssymb}
\usepackage{amsmath}
\usepackage{graphicx}
\usepackage{enumitem}
\usepackage{bbold}
\usepackage{slashed}
\usepackage{hyperref}

\newcommand\mathC{\mkern1mu\raise2.2pt\hbox{$\scriptscriptstyle|$}
        {\mkern-7mu\rm C}}              % The complex  numbers
\def\be{\begin{equation}}
\def\ee{\end{equation}}
\def\bear{\begin{eqnarray}}
\def\eear{\end{eqnarray}}

\newcommand\bra[1]{{\langle {#1}|}}
\newcommand\ket[1]{{|{#1}\rangle}}

\def\a{\alpha}
\def\b{\beta}

\def\dd{\mbox{d}}

\def\o{\omega}
\def\bra{\langle}
\def\ket{\rangle}
\def\a{\alpha}
\def\b{\beta}

\def\o{\omega}

\def\pa{\partial}

\newcommand{\sm}[1]{\mbox{\scriptsize #1}}
\newcommand{\tn}[1]{\mbox{\tiny #1}}
\renewcommand{\@}[1]{\sqrt{#1}}
\renewcommand{\le}[1]{\label{#1}\end{eqnarray}}
\newcommand{\bea}{\begin{eqnarray}}
\newcommand{\eea}{\end{eqnarray}}

\newcommand{\eq}[1]{(\ref{#1})}

\def\ffract#1#2{\raise .35 em\hbox{$\scriptstyle#1$}\kern-.25em/
\kern-.2em\lower .22 em \hbox{$\scriptstyle#2$}}

\begin{document}

\pagestyle{empty}

\centerline{{\Large \bf The Heuristic Function of Duality}}
%\vskip 0.5cm
%\centerline{{\Large\bf Equivalent Physics or New Physics?}}

\vskip 1cm

\begin{center}
{\large Sebastian De Haro}\\
\vskip 1truecm
{\it Trinity College, Cambridge, CB2 1TQ, United Kingdom}\\
{\it Department of History and Philosophy of Science, University of Cambridge\\
Free School Lane, Cambridge CB2 3RH, United Kingdom}\\

\vskip .7truecm
{\tt sd696@cam.ac.uk}
\vskip 1truecm
\today
\end{center}

\vskip 7truecm

\begin{center}
\textbf{\large \bf Abstract}
\end{center}

I conceptualise the role of dualities in quantum gravity, in terms of their functions for theory construction. I distinguish between two functions of duality in physical practice: namely, discovering and describing `equivalent physics', vs.~suggesting `new physics'.  I dub these the `theoretical' vs.~the `heuristic' functions of dualities. The distinction seems to have gone largely unnoticed in the philosophical literature: and it exists both for dualities, and for the more general relation of theoretical equivalence. The paper develops the heuristic function of dualities: illustrating how they can be used, if one has any luck, to find and formulate {\it new theories}. I also point to the different physical commitments about the theories in question that underlie these two functions. I show how a recently developed schema for dualities articulates the differences between the two functions.\\
\\

\vskip .5truecm


\newpage
\pagestyle{plain}

\tableofcontents

\newpage

\section{Introduction}\label{intro}

There is a use of duality and of theoretical equivalence that seems to have gone largely unnoticed in most of the literature, and on which this paper aims to zoom in: it is the distinction between what I shall call the theoretical vs.~the heuristic functions of both dualities and of theoretical equivalence (Section \ref{dphysics}).\footnote{The only exceptions I am aware of are: Rickles (2011, 2013, 2017), Dieks et al.~(2014), and De Haro and Butterfield (2017), in the case of duality, and Coffey (2016), in the case of theoretical equivalence. I will discuss the Rickles' views in Section \ref{ricklesd}, and Coffey's in Section \ref{incompatib} (cf.~also De Haro (2019:~\S3.4.1)). However, I here anticipate that a philosophical analysis of the heuristic function of dualities does not yet exist in the literature: and so, this paper aims to fill that gap.
\label{except}} By applying the schema for dualities from De Haro (2016) and De Haro and Butterfield (2017) to the heuristic function of duality, the paper aims to shed light on the practical use of duality and theoretical equivalence to {\it construct new theories} out of approximately dual models---which is the task of the heuristic function of dualities (Section \ref{compare}). If one is lucky, that is: for heuristics, of course, never lead mechanically, or with deductive certainty, to novel theories. In other words, I will regard dualities as {\it tools, or methodologies, for theory construction}. To do so, I will place dualities against a philosophical background discussion about the aims of scientific theories, about tools for theory construction, and about the different functions that those tools have (Section \ref{toolstc}). The discussion should  be useful for understanding different approaches to theory construction in high-energy physics and in quantum gravity more generally.\\

Dualities have become standard tools for theory construction in theoretical physics: since, at least, the discovery of position-momentum duality in quantum mechanics, and of electric-magnetic duality. But even more so, recently, especially in quantum gravity. In string theory in particular, dualities are seen as central tools for finding the still unknown M theory, as well as for learning a great deal about already existing theories.

Analysing the role of dualities as tools for theory construction in quantum gravity should help us to conceptualise some dominant practices in this field: and, in particular, it should be instrumental in explaining the importance that scientists ascribe to dualities in the overall string theory and M theory programmes. As such, a good conceptualisation of dualities, as elements of scientific practice, could also be instrumental for broader questions of theory assessment and of the progressiveness of the quantum gravity programmes---a question which I will nevertheless not take up here.\\

We can see my distinction between the theoretical and the heuristic functions of dualities or, more generally, theoretical equivalence, in a widespread difference in how they are discussed. On the one hand, we are told (sometimes with the addition of several exclamation marks!) that dualities imply that very different theoretical descriptions give rise to the same physics, or {\it equivalent physics}: and so, a theory of gravity in $D$ dimensions is dual to a gauge theory in $D-1$ dimensions (which goes under the name of `gauge-gravity duality', or `the holographic principle'). Or a theory in a very large volume, $V$, is dual to a theory in a very small volume, $1/V$, in appropriate units (this is called `T duality'). And so on. But, on the other hand, we are also told (sometimes adding even more exclamation marks!), that dualities point to {\it new physics}: and so, holographic dualities `point towards a new definition of string theory', or the dualities between the five different 10-dimensional string theories and one supergravity theory (some of which are related by the {\it same} T duality mentioned before) `point to the existence of a, so far unknown, 11-dimensional M theory'. And so on.\\

Although the philosophy of dualities is now thriving,\footnote{See, for example, a recent special issue edited by E.~Castellani and D.~Rickles (2017), in {\it Studies in History and Philosophy of Modern Physics}. For some recent philosophical discussions of dualities in string theory, see e.g.~De Haro et al.~(2015), De Haro (2016), Dieks et al.~(2014), Fraser (2017), Huggett (2017), Read (2016), Rickles (2011, 2013, 2017), Matsubara (2013).} the recent philosophical literature has---apart from the occasional mention---failed to analyse in detail this second aspect: i.e.~physicists' claims that dualities `point to new physics'. 
The literature also does not seem to have noticed the different kinds of claims that physicists make about dualities, and the kinds of expectations that are associated with such claims. In fact, the recent philosophical analysis of dualities has almost unanimously sided with the former interpretation of dualities: as cases of theoretical (and sometimes also physical) equivalence. This is understandable for the still young literature on dualities: after all, there is a venerable tradition of equivalence in the philosophy of science, which is rooted in logic and mathematics---and philosophers are less prone to mingle with physics that is not settled, or with theories that are still under construction. Thus `heuristics' is often left to the physicists. 

The general philosophy of science literature on theoretical equivalence, on the other hand, {\it has} noticed---and discussed in some depth---a similar distinction. But it is only similar, and not the same; because in the general context of theoretical equivalence, the issues that have been discussed so far are slightly different. Coffey (2016:~Sections 3.1-3.2) has noticed an `asymmetric treatment' of theoretically equivalent cases, which is similar to the distinction I am drawing here. But his treatment differs widely from mine, as I will discuss in Section \ref{incompatib}.

Thus, by siding with the `equivalence' account of duality and its corresponding use, which---I will agree---is indeed the correct account if one wishes to explicate the {\it nature} of duality,\footnote{For a detailed discussion of the relation between duality and theoretical equivalence, see De Haro (2019a).} the philosophical literature has left unanalysed (and I will substantiate this claim in Section \ref{ricklesd}) the main use that physicists make of dualities: namely the construction of {\it new} theories, as in for example the influential M theory programme. \\ %In fact,  the recent philosophical literature on duality, having not analysed this {\it heuristic function} nor even identified the tension has rendered the heuristic function almost {\it incomprehensible}.\\ 

Thus: the distinction between the theoretical and the heuristic functions is this: on the one hand, dualities describe {\it equivalent} theories (i.e.~they make new connexions between the physics described by different-looking, but {\it given}, theories, e.g.~by describing the common core that is shared between two theories). They assume we have almost complete control over those theories, so that duality conjectures can be used to develop a theory that describes the common core. And on the other, dualities are used to develop {\it new theories}, which apparently go {\it beyond} that common core, which is supposed to be the theory.\footnote{For an explication of dualities, see De Haro (2016, 2019a) and De Haro and Butterfield (2017).} 

One might think that there is an apparent tension here: if duality only expresses the equivalence between already existing theories, then it is not entirely clear how duality can help develop a theory that {\it supersedes} the two existing theories, or develop a candidate theory that will succeed them---thereby invalidating the preceding theories, and their duality relation. But I will argue that the tension is indeed only apparent: it corresponds to two different {\it functions} of duality in scientific practice. Namely, duality-as-theoretical-equivalence assumes that the two theories are well-defined, and requires that their theoretical descriptions be exactly equivalent; while the accounts that we are given, for how new physics arises from dualities, invariably assume that both the duality and the theories involved are {\it not exact}, and in fact {\it cannot} be rendered exact in the context of the theory yet to be developed, which only instantiates duality {\it approximately}. In fact, the physics literature sometimes moves seamlessly back and forth between these two views of duality: and only confusion can ensue from the mixing of these two functions.

In this paper, I will use the Schema for duality from De Haro (2016) and De Haro and Butterfield (2017) to clarify the distinction between these two different functions and to develop in detail the heuristic function. As such, the paper can be seen as an application of that Schema, thus further supporting the Schema's applicability.

The plan of the paper is as follows. In Section \ref{schema}, I give the details about the Schema for duality, from De Haro (2016) and De Haro and Butterfield (2017), that I will use in the rest of the paper. In Section \ref{toolstc}, I give some background about the idea of tools for theory construction, and define the two functions of these tools that I will consider. In Section \ref{dphysics}, I expound the basic distinction between the two functions of duality. In Section \ref{compare}, I expound the heuristic function of dualities. In Section \ref{compareo}, I compare the Schema to other recent philosophical work on dualities. Section \ref{conclusion} concludes.

\section{The Schema for Duality}\label{schema}

In this Section, I summarise the Schema for duality, from De Haro (2016) and De Haro and Butterfield (2017), which we will use in later Sections. In Section \ref{thmod}, I introduce theories and models, and in Section \ref{0dual} I give the Schema's conception of duality.

\subsection{Theories and models}\label{thmod}

The core notion of the Schema is that of a {\it bare theory}: an uninterpreted, abstract mathematical structure with a set of rules for forming propositions, i.e.~an abstract calculus. A bare theory could consist of a set of axioms or a set of equations. But, to be specific, the Schema considers a bare theory as a triple, $T:=\bra{\cal S},{\cal Q},{\cal D}\ket$, of a structured state space, ${\cal S}$, a structured set of quantities, ${\cal Q}$, and a dynamics, ${\cal D}$, consistent with the relevant structure. `Structure' here refers: first, to symmetries which may act on the states and-or the quantities, e.g.~as automorphisms of the state space, $a:{\cal S}\rightarrow{\cal S}$. And second, `structure' also refers to the set of rules for inferring propositions, i.e.~for assigning values to the quantities (e.g.~as a map ${\cal Q}\times {\cal S}\rightarrow\mathbb{R}$).

This conception of theory is as yet: 

(i) {\it uninterpreted}, i.e.~there are no rules for interpreting elements of the theory as quantities `in the world',

(ii) {\it abstract}, i.e.~there are rules for forming propositions that assign various values to elements of the triple, but there are no (there need not be) rules for practical {\it computation}. The latter typically require further definitions.

Nevertheless, this conception is {\it physical}: for choosing a set of states, quantities, and a dynamics as one's theory, even if still uninterpreted, is a physical choice which constraints the descriptive capacities of the theory (in particular, it constrains the ``number of physical degrees of freedom'' of the theory).\\

The task of (i), i.e.~interpretation, is done by the interpretation map(s), as follows. An {\it interpretation} is a set of partial maps, preserving appropriate structure, from the theory to the world. The interpretation fixes the reference of the terms in the theory. More precisely, an intepretation maps the theory, $T$, to a domain of application, $D_W$, within a possible world, $W$, i.e.~it maps $I:T\rightarrow D_W$. Using different interpretation maps, the same theory can describe different domains of the world, and even different possible worlds. For more details, also about the kinds of maps required, see De Haro (2016:~\S1.1.2).\\

The task of (ii), i.e.~of providing structure for computation, is done by the theory's models:

A {\it model} $M$ of a bare theory $T$ is a realization, or mathematical instantiation: i.e.~it is a mathematical entity having the same structure as the theory, and usually some specific structure of its own. We will use a more specific notion of model, as a representation of the theory, i.e.~a homomorphism from the theory to some other known structure: but not a merely mathematical representation, for we will make the following physical distinction within the homomorphism. A model of a physical theory naturally suggests the notion of, on the one hand, the {\it model root}: the realization of the theory, usually its homomorphic copy; on the other, the {\it specific structure}: that structure which goes into building the model root, and is not part of the theory's structure (i.e.~not part of what the theory regards as physical), and which gives the model its specificity. Part of this structure is normally used for calculations within the model---but calculations can of course be done in different ways, using different specific structure. 

It is helpful to have a schematic notation for models that exhibits how to augment the structure of a theory with specific structure: 
\bea\label{MnM}
M=\bra m,\bar M\ket~.
\eea
Here, $m$ is the model root, and $\bar M$ is the specific structure which goes into building $m$. In cases where $T$ is a triple, $m$ must itself be a triple with properties that are homomorphic to those of $T$. We call it the {\it model triple}. Thus the model $M$, Eq.~\eq{MnM}, is a {\it quadruple}: with the triple $m$ containing the states, quantities, and dynamics, and $\bar M$ the specific structure that goes into constructing the model triple. Like bare theories, models (and model triples) are, at this stage, uninterpreted (and an interpretation can again be added as a set of partial maps). Thus, the distinction between dealing with a purely mathematical representation of a theory, and dealing with a physical representation, is in the fact that the model makes a distinction between what is physical, by the lights of the model (the homomorphic copy of the theory) and the specific structure ({\it anything else} that the model may contain: such as an auxiliary calculus).\\

This usage of `model' diverges from the normal usage of a model as a particular solution of a theory (or a particular trajectory in the space of states). What we here call a `model' is often called a `theory'. This usage is motivated by dualities: dualities relate different `theories', as being different formulations of `one underlying theory': and so, they suggest that we should push the usage of both `theory' and `model' ``one level up'', while maintaining their mutual relation. In view of this basic fact about dualities---what were seemingly different theories are now just one theory---we are led to allow a more general notion of theory, and of model.

\subsection{The conception of duality}\label{0dual}

Having given conceptions of theories and of models, we give, in this Section, the conception of duality:

A {\it duality} is now defined as an isomorphism between model roots of a single bare theory, where the model roots are taken to be representations of the theory. Theories may have many representations: representations that are isomorphic to the original theory and representations that are not isomorphic to the original theory. But if we have two (or more) representations that are isomorphic to each other (whether they are isomorphic to the original theory or not), we have a duality. 

More specifically, under the conception of a bare theory as a triple from Section \ref{thmod}, a duality is an isomorphism of model roots, $m_i$, which are model triples. That is, given a set of models, $M_i=\bra m_i,\bar M_i\ket_{i\in I}$ (where $I$ is an index set that labels the different models), notice that each model root can be written as a triple, i.e.~$m_i=\bra {\cal S}_i,{\cal Q}_i,{\cal D}_i\ket$. Here, ${\cal S}_i$ is the set of states for the model $M_i$, ${\cal Q}_i$ its set of quantities, and ${\cal D}_i$ its dynamics. Duality between two such models, $m_i\cong m_j$ for some $i,j\in I$, now comes down to a triple of isomorphisms, one for each of the three items in the triple.\footnote{In most cases, choosing a dynamics is singling out a quantity, in the set ${\cal Q}$, as the Lagrangian of Hamiltonian ``driving'' the dynamics. In this case, the third isomorphism condition is not an independent map, but it is a meshing condition between the choices of dynamics, for the two dual models, and the duality map: i.e.~it is an equivariance condition between the two duality maps and the dynamics. See De Haro and Butterfield (2017:~\S3.2.1) for more details.}

\section{Tools for Theory Construction}\label{toolstc}

This paper develops the heuristic function of dualities in constructing new theories of quantum gravity. Our topic is thus {\it theory construction}, and in Section \ref{dphysics} I will argue that duality is one of the tools, or methodologies, which are available for theory construction.\footnote{The word `tool' is standardly used in the literature on scientific understanding on which I build (De Haro and De Regt (2017) and references therein), and so I will stick to its use.} Therefore, we should have a basic sense of how tools are used in scientific theories, and of how they contribute to the aims of scientific theories. In Section \ref{aimssc}, I discuss different aims of science. In Sections \ref{thef} and \ref{heurf}, I discuss the theoretical and the heuristic functions of the tools, respectively. 

There are two things that my discussion in this Section will {\it not} attempt to do: (i) to give a complete list of tools (whatever that would be), (ii) to give analytic definitions of the different tools. As noted in De Haro and De Regt (2017) for tools for scientific understanding (and it seems to be the same in this case): such an aim would be illusory. Thus my aim will be much more modest: I will view tools as `just tools towards achieving a goal, and not as necessary or as sufficient conditions for achieving a goal'.\footnote{See e.g.~Laudan (1984:~p.~44): `cognitive values [which elsewhere, on p.~xii, he also calls `aims'] underdetermine methodological rules' (which I call `tools'). Though my terminology differs from Laudan's, I agree that there is no `covariance' between the methods and the aims of a theory. In this sense, I also agree with the general gist of his `reticulated model' (pp.~62-63): viz.~that theories, methods, and aims are not related linearly or strictly hierarchically, but are intertwined in relations of mutual dependency.} On this practical approach, one admits that goals can be achieved by many different means: and that classifications of tools, even if useful, will therefore always remain incomplete. And one also admits that tools can be used in different ways, and that different tools can be used to fulfill the same function: therefore, for most functions we wish a tool to fulfill, there will always be an element of specificity, ad-hocness, or even vagueness.\footnote{In the philosophy of science, Laudan speaks about underdetermination of methodological rules [tools] by aims (1984:~p.~44). This, however is still very general. I will be primarily concerned with the {\it function} which a tool can fulfill. In the philosophy of technology, multiple function ascription is regarded as a virtue that it is good for a theory to have. See Houkes and Vermaas (2010:~p.~59).}

That said, there are, of course, substantive philosophical questions about how tools are used, and about how they achieve their aims. 

{\it Theory construction vs.~the context of discovery.} My talk here of `theory construction' is not meant to imply my endorsing a sharp philosophical distinction between what has traditionally been called the `context of discovery' vs.~the `context of justification': about which one can indeed have justified reservations (see e.g.~Radder (1991:~pp.~222-223)). Rather, there is, of course, a {\it scientific activity} of theory construction---which is what the current programme of quantum gravity is largely involved in. As such, theory construction entails both the discovery and the development of new theories, as well as their justification and assessment. In this paper, I aspire to give an anatomy of the use of dualities, within the broader picture of philosophers' thinking about heuristics.

\subsection{Aims of scientific theories}\label{aimssc}

In this Section, I discuss some of the aims of theoretical enquiry, i.e.~the aims of scientific theories, and how these can sometimes lead to tensions between the different {\it functions} of the elements comprising a theory (viz.~bare theory, interpretation, models).\footnote{For more details on my use of these notions, see De Haro (2016) and De Haro and Butterfield (2017).} I will argue that some of these tensions are substantive; although they do not necessarily lead to contradictions and they can be resolved, depending on a variety of factors. The aim here is to provide the background against which, in Section \ref{dphysics}, I will state the distinction between the two functions of duality.

A theory can be used to {\it describe} the world in detail and accurately, on a suitable interpretation. Other uses of theories are {\it instrumental}: for example, using the theory as a calculational tool to get ``quick and dirty'' results about a situation of interest, without paying attention to other contextual details that are irrelevant for, say, the aim of quantitative {\it prediction} in a specific situation---though prediction need, of course, not always be instrumental. These two uses---the descriptive and the instrumental---are of course tuned to corresponding aims of theories: and so there is no real incompatibility here. Debate can then ensue about how to {\it prioritize} those goals, about how the goals are related, or about the conditions under which a given goal is worth pursuing, but not about whether a theory can in fact {\it have} those two different uses: for they are both legitimate.\footnote{That scientific theories can have `several different, even mutually incompatible, goals'  (Laudan (1984:~p. 63)), should be uncontroversial. `There is no single ``right'' goal for inquiry because it is evidently legitimate to engage in inquiry for a wide variety of reasons and with a wide variety of purposes'' (Laudan (1984:~pp.~63-64)). See also Toulmin (1961:~\S2).}

Furthermore, a theory, and especially its {\it interpretation}, can also be aimed at explanation or at understanding (Toulmin (1961:~\S2), De Haro and De Regt (2018:~\S1.1), De Regt (2017)). For example, Ruetsche (2011:~pp.~3-4) has contrasted an `ideal of pristine interpretation', which sees the business of interpretation as a `lofty' affair that is only concerned with the general question of which worlds are possible according to the theory, and not with the application of the theory to actual systems. She argues for `a less principled and more pragmatic approach to interpreting physical theories, one which allows `geographical' considerations to influence theoretical content, and also allows the same theory to receive different interpretations in different contexts.' (p.~4).

A similar difference is sometimes seen between differing uses of the word `model': while the philosophy of physics literature tends to endorse the semantic conception of models, i.e.~as the set of worlds that are possible according to the theory, in the general philosophy of science the notion of models involves both context and approximation---there, models mediate between the theory and concrete phenomena, which obtain under definite circumstances.\footnote{My use of `model' in Section \ref{thmod} differs from both of these uses. Although my use of `model' in an essentially model-theoretic sense, what I here call `model' is often called a `theory', as I mentioned in Section \ref{thmod}. So I will call the other construal of a model, as a description of a concrete system that mediates between theory and phenomena, a `model description' (of a system, or of a phenomenon).}

One must of course recognise that behind these disagreements in the literature, about what interpretations and models `really are', there can be---and there often are---larger philosophical differences: between theoretical and practical approaches to science, between realist and anti-realist positions, %between reductionist and anti-reductionist positions, 
or between trust in the notion of laws of nature vs.~belief in a `dappled world'---just to mention some.

But one must also admit that, these larger differences aside, there is a clear way to go about resolving the disagreements about the philosophical notions of interpretation and model, i.e.~about the notions which comprise a theory: namely, by recognising that they are tuned to diverse, but equally legitimate, aims for which the notions involved (viz.~bare theory, interpretation, model: cf.~Section \ref{thmod}) are used. Thus for example, the question of whether a model is a possible world in which the theory is true, or is a contextual and specific application of a theory to describe a phenomenon, 
%is to some extent verbal: since there is no monopoly on the word, both construals of it are legitimate, each of them being tuned 
will receive different answers depending, to a large extent, on the kind of question one is asking. Indeed, it will depend on the specific purpose one wishes the word `model' to fulfill---that of, say, expounding the descriptive capacities of a theory vs.~that of expounding its applicability to specific cases. Again, the two uses are legitimate, though they will no doubt lead to different philosophical accounts. \\

Like theories, dualities and relations of theoretical equivalence can also have several aims: not only in the construction of new theories, but they can also have instrumental uses, and they can be used to explain or to attain understanding (De Haro and De Regt (2017~\S2,3.1)). 

Here, I will concentrate on theory construction. But even within the general aim of theory construction: a duality, or a relation of theoretical equivalence, can have different functions. That is, there are different ways in which a theory can be constructed, using dualities or equivalences. My point (in the next two Sections) will be that the two functions---theoretical and heuristic---should not be confused, because they lead to different results. Indeed, there are two functions of dualities which:

(i)~~~Are both expressed in the physics literature.

(ii)~~\,Express actual scientific disagreements about what the unifying theory underlying a duality relation looks like (e.g.~M theory). 

(iii)~~Do in fact correspond to different aims, and different uses, of the notion of duality, and lead to different results.

(iv)~~Nevertheless, do not lead to contradictions; i.e.~they exemplify different ways in which theory construction can be approached.

%As I mentioned in the Introduction, the philosophical literature has not noticed the existence of these two functions, nor their incompatibility. Thus, it has only latched on to one of the uses of duality made in the physics literature, and has left its other uses unanalysed. %Hence there is something to be explicated here.

\subsection{The theoretical function}\label{thef}

In this Section, I briefly introduce the theoretical function, that tools for theory construction can have.

There are of course many kinds of theoretical tools used in physics: for example symmetry arguments, analogies, approximative relations, and indeed dualities. These tools can take on different functions, i.e.~they can be put to use in different ways, under different constraints, and for different purposes (even if the general aim, as we assumed throughout this Section, is invariably taken to be `theory construction').\footnote{There is a lively debate, in the philosophy of technology, about the correct notion of `function' (see Houkes and Vermaas (2010:~Chapter 3)). I will not need to develop a function theory, although I believe that the analogy between artefacts and scientific theories can be used to specify more precisely what one means by a `function' of a scientific theory.\label{functionpt}} The theoretical function I have in mind is aimed at developing a {\it given} theory, i.e.~not a novel theory, according to constraints. It is the aim of extracting the content of a theory ``that is somehow already there'', even if only implicitly, using a set of rules. The set of rules is then the tool in question, though use of the tool of course never by itself guarantees success.

For example, once one knows the Hamiltonian (i.e.~the energy function) describing a system in classical mechanics, one can use Hamilton's variational principle to derive the equations of motion for that system. In doing so, one may encounter problems (e.g.~the difficulty of how to choose appropriate boundary conditions for a given situation), so success is not guaranteed. Nevertheless, the equations of motion are, in essence, ``already there'' once the Hamiltonian is given, since there is a set of steps which lead from a Hamiltonian to the equations of motion, partly deductively. That set of steps, i.e.~Hamilton's variational principle, is the tool in question, used in its theoretical function of finding the equations of motion, i.e.~of extracting the full theory (of which the equations of motion constitute the dynamics). 

The theoretical function, as just presented, comes with a partly deductive procedure to get the theory, $T$: and so, the theory is, in a way, already there from the start.\footnote{A theory, $T$, thus formulated may of course have unforeseen consequences, such as the existence of new (non-isomorphic models). Elucidating the physical interpretation of those new models will be the task of the heuristic function. Notice that the theoretical function of course has an element of heuristics of its own (cf.~the remark above about boundary conditions): but the method of the theoretical function is a largely {\it deductive}---sometimes even a mechanical---one, i.e.~it follows prescribed rules.} Thus the method is not aimed at finding physical novelty: even if in some cases it may find some novelty---precisely in those cases in which complete deduction fails. Rather, it is aimed at making more perspicuous the conceptual and-or mathematical presentation of the theory $T$. This is what I call the theoretical function of a tool.

One might object: why call duality a `tool', and the use made of this tool a `function'? Is it not simply a case of providing a mathematical proof of duality, i.e.~is duality not the thing we wish to prove, rather than the tool to achieve a goal? 

The answer is No, and for tree reasons. First: a proof of duality, given a set of models, only requires the {\it existence} of a theory, of which the models in question are representations, and of an isomorphism between the models: the proof does not require the actual {\it construction} of the theory, $T$, which is the aim of the theoretical function (cf.~the definition of duality in Section \ref{0dual}, does not explicitly mention the theory, but only its models). Thus the aim of {\it constructing} a theory, $T$, is more ambitious than the aim of proving duality. 

Second, the theory thus constructed need not be unique, as emphasised in De Haro and Butterfield (2017:~\S 2.4). Thus, there is judgment involved in deciding which theory, $T$, is the most appropriate one, in a specific situation. 

Third, there is also choice that scientists can make between attempting to try to construct the common core, i.e.~the theory $T$, or not constructing it. For the construction of theories exhibiting a duality is not a necessary aim of scientific theories: a theory with a duality need not necessarily be better than a theory without it. In other words, theorists faced with a duality are free to construct the theory $T$ or to not construct it, and there can even be a choice of the theory $T$ among a number of competitors. Thus, duality is a tool towards theory construction which scientists can choose to try to construct such a theory, as a common core, if they so wish: and the function duality then has is a theoretical one. 

An analogy with symmetries and with approximative relations is helpful here. Imagine a rule that produces, given an input state, an output set of states, according to a symmetry principle (e.g.~given a wave-function with given energy, symmetry considerations are used to produce other wave-functions with the same energy). This is done by the theoretical function.

But now imagine a more adventurous use of symmetry in which, given an equation of motion that does not display a given symmetry (e.g.~because it is written in a specific gauge or coordinate system), one writes the equation in a manifestly symmetric way (e.g.~in a gauge-invariant way, or as a covariant equation). Once again, one can here give rules for such a procedure, of symmetrisation, or covariantization. This use of symmetry also falls under the theoretical function, for two reasons: 

(i) there is a well-defined general rule, saying `for any equation A of a certain kind that does not exhibit a symmetry, there is an equation B that is manifestly symmetric', 

(ii) it does so in such a way that the number of degrees of freedom is not modified, in particular it is required that no new physical degrees of freedom are introduced. 

Thus, though not strictly deductive in the logical sense, this use of the theoretical function still operates according to a general rule, and it does not introduce ``new physics''.

Similarly for an approximative scheme: this theoretical `function' is a matter of a reduction, or linkage, between two theories.\footnote{Philosophers' traditional account of reduction is that of Nagel (1961:~pp.~351-363). For the notion of approximation which I use, see De Haro (2019).} Given a basic, or `bottom', theory $T_{\sm b}$, the approximation scheme is a rule that produces a new, `top', theory $T_{\sm t}$. Again, though the success of the application of this rule is not guaranteed: if it succeeds, then one ends up with a theory $T_{\sm t}$ that is obtained from $T_{\sm b}$. Because there is reduction, or at least linkage, there is a sense in which the degrees of freedom of $T_{\sm t}$ can be taken to be derived from those of $T_{\sm b}$ (under suitable assumptions about the approximation).\footnote{I am assuming that there is no novelty in the interpretation of the top theory, relative to the basic theory (for an account of ontological and epistemic novelty and the associated notions of emergence, see De Haro (2019)). For I take a procedure that helps develop a {\it new interpretation} of a theory to be part of a heuristic function as well. The example of approximation here is what Radder (1991) has called `correspondence from L [$T_{\sm b}$] to S [$T_{\sm t}$]. It is the kind of heuristics considered by Post, Krajewski, and Fadner (mentioned in Radder (1991)). With Radder, and contra these authors, I do not take heuristics to be applicable only, or even mainly, at the formal level. }

\subsection{The heuristic function}\label{heurf}

In this Section, I briefly introduce the heuristic function that tools for theory construction can have.

Whewell (1860:~p.~480) described `heuristic' as the `art of discovery', which, he admitted, was not to be understood as `a kind of Logic'. A narrower conception of heuristics is as a set of `efficient rules or procedures for converting complex problems into simpler ones' (Hey (2016:~p.~472)). It is the former conception which I have in mind in this paper: a tool as used in the art of discovery, and in the construction of new theories.\footnote{The heuristic function is of course not limited to the construction of new theories. It in fact plays an important role as well in the application of known theories, in theory confirmation, etc. However, as mentioned before, I will here restrict the discussion to theory construction, which is relevant to Section \ref{compare}'s main example, the quantum gravity programme.} It is a {\it tool}, rather than a {\it rule}, because success in theory construction is never guaranteed; nor can the tool be applied mechanically, as the phrase `efficient rule' would suggest. 

Indeed, whenever general, and more or less mechanical, {\it rules} are involved in theory construction, I will take the corresponding use of the tool to belong to the theoretical function, as discussed in Section \ref{thef}, rather than to the heuristic function (assuming that the rules also do not introduce new degrees of freedom or interpretative novelty). The heuristic use of a tool involves craftmanship and creativity, and should lead to the formulation of {\it new theories}, which contain new physics: rather than merely reformulating more precisely, or more perspicuously, already (implicitly) known theories, according to systematic rules. 

The heuristic function will obviously have some rules of its own (having to do with the sorts of constraints that the new theories should satisfy), but the defining mark lies in the theoretical and physical novelties that are its aims. Novelty in the theory's formalism includes: the number and nature of states and quantities (or `physical degrees of freedom'), the dynamics, and the rules for calculating physical quantities (cf.~Section \ref{thmod}). Novelty in the interpretation is novelty in the theory's reference to worldly items, which includes cases of ontological emergence.

Let me illustrate this in the examples given in the previous Section. There, we considered a system for which the Hamiltonian was given. The theoretical function then was a rule that gave us the equations of motion for this system. Now consider a system for which the Hamiltonian needs to be found. The main difference between the two cases is that, in the former, there is a partly deductive procedure. In the second case, there is no such procedure, and the arguments required are of a different kind. Scientists indeed use heuristics when, given a physical system, they try to find a Hamiltonian describing this system. The procedure in question may involve the writing down of parts of a Hamiltonian (or limits of it) which they already know from similar systems: but it also involves educated guesses about those parts of the Hamiltonian which they do not yet know, e.g.~because they describe some of the system's novel, or even unique, features. Such tentative guesses are usually informed by different kinds of arguments: symmetry arguments, combined with arguments about the number and kind of degrees of freedom to be described, assumptions or constraints about the admissible kinds of interactions, etc. But even if physicists are able to come up with fairly systematic rules constraining the admissible classes of Hamiltonians (though this usually only works for a class of {\it similar problems}), in the end there is no mechanical, or indeed general, rule for writing down the Hamiltonian describing a physical system: it is ultimately always a matter of creativity, craftmanship, and some luck, and the best one can do is verify that it describes the target system accurately, in specific situations. 

Recall the example, at the end of Section \ref{thef}, of an approximative scheme (e.g.~$\hbar$ small compared to the action) relating the top theory, $T_{\sm t}$, to the bottom theory, $T_{\sm b}$. The heuristic relationship between $T_{\sm b}$ and $T_{\sm t}$ now goes in the opposite direction to that discussed in Section \ref{thef}. Given an approximative theory, $T_{\sm t}$, and given an approximation scheme from which one believes $T_{\sm t}$ (or a theory close to it) is obtained, physicists' job is now to try and guess, or to somehow reconstruct, the basic theory, $T_{\sm b}$. Again, such educated guesses are subject to constraints, but $T_{\sm b}$ can ultimately only be justified if it describes more phenomena than $T_{\sm t}$ does.\footnote{There is of course no claim here that $T_{\sm b}$ is descriptively strictly better than $T_{\sm t}$, in each and every respect. Kuhn losses may well be our fate! But at least $T_{\sm b}$ does explain the success of $T_{\sm t}$. This possibility of a `loss', when going from $T_{\sm t}$ to $T_{\sm b}$, is of course the counterpart of emergence, in the reverse direction. See De Haro (2019).\label{thloss}}

\section{The Two Functions of Duality}\label{dphysics}

In the previous Section, I discussed two of the functions that tools for theory construction can have: a theoretical function and a heuristic function. In this Section, I will discuss those two functions for dualities in string theory, and show how they differ. In \S\ref{bast}, I expound the basic distinction, using quotations from the physics literature. In \S\ref{incompatib}, I argue that there is only an apparent tension: not an incompatibility.

\subsection{The distinction in string theory}\label{bast}

In this subsection, I describe how the distinction between the theoretical and the heuristic functions plays out in string theory. To this end, I use quotes from the physics literature (since, as I mentioned, the {\it philosophical} literature on dualities has not identified this tension).

After introducing, in Section \ref{stringMth}, the string and M theory programmes, I proceed in two steps. In Section \ref{exacteq}, I describe the {\it theoretical function} of duality. This is the function which the philosophical literature has focused on. I have discussed this theoretical function in De Haro (2019a), to which I refer for details. Then, in Section \ref{dualheur}, I argue that there is a second, heuristic, function, which duality plays: that second function is certainly no less important than the theoretical function, and so it deserves philosophical scrutiny. %I will argue that the two functions are incompatible in some ways, but not in others. 

\subsubsection{Motivating duality: string theory and the M theory programe}\label{stringMth}

I first briefly introduce, in this Section, the main ideas behind the string theory and M theory programme: and, in particular, the role of duality within that programme.

String theory is a candidate theory for the unification of general relativity and quantum field theory. Its basic assumption is that matter is made of strings, i.e.~extended, one-dimensional objects that can vibrate, move around in spacetime, and interact by joining or by splitting. 

For string theory to be mathematically consistent, 10 spacetime dimensions are required for the strings to move in (6 of which are thought to be curled up, so that they are inaccessible to current experiments). In the low-energy limit, string theory is well-approximated by supergravity theories, i.e.~supersymmetric extensions of Einstein's theory of general relativity, which are also 10-dimensional, and compactified down to four dimensions.

%In the first half of the 1990s, it became clear that the theory, as known, was only part of a much richer structure. In 1995, Polchinski discovered new objects within the theory, called D-branes. These are higher-dimensional objects which in some sense generalise the string (they have an extension in two or higher spatial dimensions, whereas the string only has one). In another sense they are different from the string, because they are not fundamental objects of the theory as given (which contains only strings) but rather boundary conditions for the string: they are higher-dimensional surfaces on which strings can end.

%D-branes were known in supergravity theories prior to 1995, but their discovery in string theory led to a revolution now know as the `sfcecond string revolution'. Witten (1995) con

Initially, five different string theories were known, differing over the precise details defining the strings. However, significant dualities were found relating them to one another. T duality, for example, relates one type of string theory on a circle of radius $R$, to another type of string theory on a circle of radius $1/R$. And electric-magnetic duality (so-called S duality) relates some other string theories.

In 1995, Witten conjectured that the five known string theories, plus in addition a sixth known, 11-dimensional, supergravity theory, were all different limits of (approximations to) a single 11-dimensional theory, which he dubbed M theory. Witten assumed the eleventh dimension to be a circle, which could be of one of two kinds. He identified the radius of this circle with the coupling constant ruling the joining and splitting interactions of the strings. For a small circle of the first kind, the string coupling is weak, so that one of the five known 10-dimensional string descriptions describing weakly-coupled strings (the so-called {\it perturbative string theory}), is accurate. For a small circle of the second kind, another of the five known versions of perturbative string theory is accurate. The other three string theories are related to these two by T and S dualities.

But, at strong coupling, the eleventh dimension opens up, and the perturbative string descriptions are no longer valid. Eleven-dimensional supergravity provides a semi-classical description in 11 dimensions, valid at strong string coupling but only as long as the length of the fundamental string is small, i.e.~in the point-particle limit of the string (or whatever replaces it in eleven dimensions). The challenge is then to find a theory valid away from the point-particle limit: this should be the sought-for M theory.

Since Witten's conjecture, two main approaches to M theory have been taken. The first is the conjecture by Banks, Fischler, Shenker, and Susskind (1997) that M theory is a theory of matrices, with eleven-dimensional supergravity as its low-energy limit. 

The second main approach is AdS/CFT, which is a series of conjectured dualities between string theory or M theory in asymptotically anti-de Sitter space (AdS, i.e.~a manifold of negative curvature), and a specific quantum field theory at the boundary of this space (where CFT stands for `conformal field theory'). Compactifying M theory on e.g.~an internal seven-dimensional manifold of positive curvature, the remaining four dimensions have negative curvature (``they are AdS''), and are dual to a three-dimensional CFT, for which exact treatments exist. This approach is more generally called `gauge-gravity duality', because it relates a theory of gravity to a quantum field theory with gauge symmetry.\\

Details aside, M theory is the main unifying conjecture behind the various versions of string theory, and dualities play a key role in the attempt to formulate M theory. What remains unclear is the precise status that dualities are supposed to have in M theory, once a non-perturbative version for it is found. Should M theory exhibit duality, or should dualities be superseded by the final theory---are they merely ``ways towards the formulation of a new theory''? 

This question is, of course, not about trying to peek into the future of theories that do not yet exist, but about the heuristic paths of investigation that one may reasonably take dualities to suggest. We will explore the role of dualities within this programme in Sections \ref{exacteq} and \ref{dualheur}. Here I anticipate by saying that the answer to this question will come down to a different function of duality.

The conjectural status of most dualities in string theory, and of M theory itself, should not be a reason to dismiss the programme as philosophically irrelevant, or as mere speculation. There are four reasons for this, which I here list:\\

First, the programme is very influential in physics: and, in the last thirty years or so, it has spawned a large number of new ideas and technical developments which (arguably) no other research programme in high-energy physics has been able to produce. Second, and more importantly, being conjectural does not mean being physically and mathematically unmotivated. For the evidence that is available for some of the string theory dualities is strong and compelling. Third, there are also rigorous results, at various levels of mathematical and physical rigour: especially about the conformal field theories, random matrix models, and quantum field theories involved, fairly rigorous mathematical results exist. Finally, it is of course simply false that philosophy should limit itself to studying theories that are already in final form and that are mathematically completely rigorous: for not only would philosophers then quickly run out of a job, but also because it is their task to clarify and assess whatever fragments of theory are available (cf.~Huggett and W\"uthrich (2013:~p.~284)). This is especially true in areas of research such as quantum gravity, where direct observations are so far absent, and so the main guidance is the---apparently very strong---requirement that general relativity and quantum field theory should be reproduced in suitable approximations, and in addition one has the requirement of mathematical consistency and the tools of conceptual analysis at one's disposal (besides what little available evidence there is from experiments and analogue experiments). Rather than making the quantum gravity attempts uninteresting for philosophers, these four reasons make philosophy relevant, even indispensable, to the programme of quantum gravity.

\subsubsection{Duality as exact equivalence: duality's theoretical function}\label{exacteq}

In this Section, I discuss within string theory the theoretical function of duality, in the sense of Section \ref{thef}, where duality is construed as in the Schema from Section \ref{schema}.

The physics literature construes duality as an isomorphism between models. This isomorphism relates the common core that the two models deem physical (i.e.~the triple of states, quantities, and dynamics). As such, duality is a formal notion, i.e.~a definite relationship between uninterpreted, but physical, models: it is a special case of theoretical equivalence. It relates triples of states, quantities, and dynamics on the two sides, preserving the structure of the models (including the values of the quantities, evaluated on the states). %While the elements of the triple are uninterpreted, such a choice of the number and nature of the degrees of freedom (i.e.~the elements of the triple) constrains the descriptive capacities of the theory. 
Thus duality is not {\it merely} a formal relation, because it deals with {\it physical} models, but by itself it makes no reference to interpretation---the latter is the question of what I will call `physical equivalence'. 

Both physicists and philosophers tend to construe duality this way. Therefore, the theoretical function of dualities, i.e.~the function that follows from the {\it nature} of duality, as outlined in Section \ref{thef}, is to establish theoretical relationships (more specifically: to establish a theoretical equivalence, as a specific kind of isomorphism) between models. These relationships typically entail relating states and quantities in one model, to states and quantities in another model, and also relating the dynamics of one model to the {\it different}, but isomorphic, dynamics of the other.

Thus dualities are very strong relationships between two models, since they relate everything that the models deem physical (namely, the model root $m$ that is within the model $M$ in Eq.~\eq{MnM}). Establishing a duality between two models thus presupposes precise knowledge of the elements of the two models (the sets of all the states and quantities, and the complete dynamics), as well as knowledge of the relations in which these elements stand (i.e.~there are not only bijections between each of the elements of the triples of the two models, but all physical structure must also be preserved). Thus establishing a duality requires a formulation of a model that captures all of those details, even if perhaps only implicitly. Full transparency of the model, or full understanding of it or perfect computational power, are of course not (and cannot be) required: but duality {\it does} require a formulation of the models that is as detailed as just described, within their domains of application. I will say that such a model (i.e.~one where all the states and quantities, and the complete dynamics, as well as the complete rules for calculations, are known and are consistent, within the domain of application of the model) is {\it exact}. %, and within the theory's limits of accuracy. %: it must describe {\it all} the states and quantities, and the entire dynamics, within its domain of application. 

Notice that this notion of being mathematically well-defined, within a domain of application, is much weaker than the requirement that a model gives a non-vague, good, or succesful description of the domain---the former is a formal requirement, while the latter is interpretative. 

Furthermore, when such models are given, and a duality between them exists, we say the duality is {\it exact}.\footnote{Exactness is a prerequisite of duality, because it is part of the isomorphism condition. My use of the phrase `exact duality' simply emphasises duality's exactness. It does not suggest that `inexact dualities' exist, which would be contradictory. Rather, dualities can be inexact, or approximate, in a different sense: namely, in the sense that
 they %do not give exact descriptions of the world, or in the sense that dualities 
are not instantiated by the models in an exact manner. Whenever I use the adjective `inexact' in connexion with dualities, it will be in this sense.}

Exactness can be proven for a number of significant dualities in physics. Simple examples are the Fourier transformation in elementary quantum mechanics, harmonic oscillator duality, and electric-magnetic duality in electrodynamics. For more sophisticated dualities in quantum field theory and in quantum gravity, the only case, so far as I know, in which the philosophical literature has proven a duality to be exact is the example of boson-fermion duality in two dimensions (De Haro and Butterfield (2017)), though in the physics literature there are other cases. Most dualities in string theory (T duality, gauge-gravity duality, S duality, etc.) are cases of dualities which are {\it conjectural}. Nevertheless, it is an important aspect of duality that {\it all} dualities are exact---as they must be, according to the above definition. 

The physics literature confirms the claim that dualities must be exact: i.e.~that the {\it definition} of duality entails that they are cases of exact, and not approximate, equivalence, within a domain of application.
% (and to a degree of accuracy which the theory itself may prescribe). 
Also, the physics literature confirms that duality is a case of theoretical equivalence, i.e.~of a formal, or mathematical, relationship between two physical models, as in Section \ref{0dual}. I will now substantiate this consensus some quotations from the physics literature, which also illustrate how physicists think about dualities. 

The literature quoted below of course also emphasises the following aspects: seemingly different physics and difference of description, but equivalence (or sameness) of theory; and the exactness of the duality, and of the theories involved, is also denoted as the theory's being `non-perturbative', i.e.~its formulation goes beyond, or does not require, perturbation theory.\\
\\
(A)~~In the Glossary of his textbook on string theory, Polchinski (1998, p.~367, my emphasis) defines duality as:
`the {\it equivalence} of seemingly distinct physical systems. Such an equivalence often arises when a single quantum theory has distinct classical limits.'

He describes one specific duality (T duality) as a case of sameness of theory, but difference of description: `T-duality is just a different description of the same theory' (p.~268). `[T-]duality is a symmetry not only of string perturbation theory but of the {\it exact} theory (p.~248, my emphasis).\\
\\
(B)~~In an influential paper putting forward the matrix model conjecture for the definition of M theory (mentioned in \S\ref{stringMth}), Banks et al.~(1996:~Abstract, my emphasis) also regard duality as an exact equivalence. Thus they write: `We suggest and motivate a {\it precise equivalence} between uncompactified eleven dimensional M-theory and the $N=\infty$ limit of the supersymmetric matrix quantum mechanics'. `If our conjecture is correct, this would be the first {\it nonperturbative formulation of a quantum theory} which includes gravity' (p.~2, my emphasis). And later they say:

`Our conjecture is thus that M-theory formulated in the infinite momentum frame is {\it exactly equivalent} to the $N\rightarrow\infty$ limit of the supersymmetric quantum mechanics described by the Hamiltonian (4.6). The calculation of {\it any physical quantity} in M-theory can be reduced to a calculation in matrix quantum mechanics followed by an extrapolation to large $N$.' (p.~11, my emphasis).\\
\\
(C)~~In an influential review on gauge-gravity duality (cf.~\S\ref{stringMth}), Aharony et al.~(1999:~p.~57, my emphasis) formulate duality in terms of sameness of theoretical description, or theory: `Thus, we are led to the conjecture that... %${\cal N}=4$ $\mbox{U}(N)$ 
Yang-Mills theory in 3+1 dimensions is {\it the same as (or dual to)}...
% type IIB 
superstring theory on $\mbox{AdS}_5\times S^5$'.\footnote{The words omitted in the quotation are technical details specifying the two theories involved, viz.~${\cal N}=4$, $\mbox{U}(N)$ super-Yang-Mills theory, and type IIB string theory.}

%``In its weakest form the gravity description would be valid for large $g_sN$, but the full string theory on AdS might not agree with the field theory. A not so weak form would say that the conjecture is valid even for finite $g_sN$, but only in the $N\rightarrow\infty$ limit (so that the $\a'$ corrections would agree with the field theory, but the gs corrections may not). 
They extend this conjecture to a full equivalence between string theory and gauge theory: `The strong form of the conjecture, which is the most interesting one and which we will assume here, is that {\it the two theories are exactly the same} for all values of $g_s$ and $N$ [i.e.~the string coupling constant and number of colours, respectively].' (p.~60, my emphasis).\\

The common thread is clear: these are all cases of conjectured, but {\it exact}, equivalences of the theoretical structures (sometimes, in a limit of the physical parameters that is relevant to the theories involved). This is in agreement with the Schema's definition of duality, given in Section \ref{0dual}, and it grounds the theoretical function of duality: namely, duality thus construed is a relationship between models that are already there and which were previously thought to be unrelated.

In light of the discussion in Section \ref{thef} on the theoretical function of a tool, we can now understand a conjectured duality as a help in finding more perspicacious formulations of a given model. This is for example the case when physicists use the better-known side of the duality to investigate the lesser-known side. This is akin to solving a problem (even: formulating a model description of a system) in momentum space, and then doing the Fourier transformation back to position space. This use of the Fourier transform, which is a deductive rule that by itself does not add any new degrees of freedom, is a translation of one model description to another, and so it belongs to the theoretical function. Unless the model description A was already known, the Fourier transform would be of no help in getting the model description B via duality. It is only when the model description A is already worked out, that we can find out more about the model description B, in a quasi-mechanical way, using the Fourier transform. The same remarks go through for other dualities, in this kind of use.

But notice the assumption behind string theory dualities: within the theoretical function, the duality relation itself {\it will not change}, once the two dual models are formulated to our satisfaction (i.e.~as a quadruple, involving the model root and the specific structure: see Eq.~\eq{MnM}). Rather, the search for a satisfactory formulation of two dual models is a search for two structures that stand in precisely the relation that is described by the duality conjecture. On this view, duality is not to be superseded in the theory one is aiming to construct: rather, {\it establishing duality} is the aim of the proof of the duality conjecture. The duality is to be instantiated by  the final pair of models: perhaps in a manifest and completely obvious way, on a sufficiently perspicacious formulation of them. I will call the theory, $T$, thus obtained the {\it common core theory}: for this theory contains the core stucture that the models deem physical (usually, a triple of states, quantities, and dynamics, as in Section \ref{thmod}), and this core structure is isomorphic between dual models, i.e.~it is their common core: viz.~the model root, Eq.~\eq{MnM}, of each of the models.

\subsubsection{Duality and approximation: duality as a heuristic for theory construction}\label{dualheur}

In this Section, I discuss within string theory the heuristic function of duality, in the sense of Section \ref{heurf}, and give some quotations from the physics literature supporting the existence, and even the essential role, of this function, in the recent programme of string theory and M theory. 

The physics quotations below also emphasises the lack of exactness of the theories involved (viz.~they are perturbative) and the use of dualities as heuristics for finding new unifying theories (or new formulations of old theories, describing more physics). The heuristic function, in the context of this literature, is then seen to be strongly linked with the aim of {\it unification}. The examples are as follows:\\
\\
(A)~~In a review paper about dualities, Dijkgraaf (1997:~p.~120, my emphasis) connects the approximate nature of dualities to the suggestion of the existence of new theories:  `The insight that all perturbative string theories are different expansions of one theory is now known as string duality... {\it It is one of the amazing new insights following from string duality that these “theories” are all expansions of one and the same theory around different points in the moduli space of vacua}.' 

`Expansion... around a point' should here be taken in the sense of, for example, a Taylor series expansion of a function about a particular point: which is captured by the notion of `approximation', discussed in Section \ref{heurf}. Dijkgraaf also emphasises the `perturbative' nature of the dual models, i.e.~their lack of validity beyond a certain order in such an expansion (a so-called `perturbative expansion'). Thus, Dijkgraaf's picture of dualities is one which regards models as {\it inexact}, and dualities as only approximately instantiated, i.e.~the dualities are valid only within a limited range of parameters, but are to be {\it superseded} by a better theory, namely what he calls `one and the same theory', of which the mutually dual models are expansions, i.e.~approximations.\\
\\
(B)~~In the paper in which Witten put forward the influential M theory conjecture, he wrote  (1995:~p.~2, my emphasis): `S-duality between weak and strong coupling for the heterotic string in four dimensions... really ought to be {\it a clue for a new formulation of string theory}.'

`Another motivation was to try to relate four-dimensional S-duality to statements or phemonena in more than four dimensions... {\it we are bound to learn something if we succeed}' (p.~2).

`...in this paper, we will analyze the strong coupling limit of certain string theories in certain dimensions. {\it Many of the phenomena are indeed novel, and many of them are indeed related to dualities}' (p.~2).

`Combining these statements with the much shakier relations discussed in the present paper, one would have a web of connections between the five string theories and eleven-dimensional supergravity' (p.~4).\\

These quotes by Dijkgraaf and Witten underline a related aspect of dualities: they use terms like `amazing', `new insights', `clue for a new formulation', `learn something', `novel phenomena'. The emphasis here, unlike the quotes from \S\ref{exacteq}, is not on the conjectured equivalence between already existing models: but on the {\it novelty of theory} which can arise once a duality between such models is understood. 

They also emphasise duality's pointing to `a new formulation of string theory': where I take it that `a new formulation' is more than just a `{\it re}formulation': for a new formulation contains something extra, not only in terms of the mathematical formalism, but also in terms of the physics that is associated with that formalism---as the other quotes confirm, when they talk about novelty of phenomena: `we are bound to learn something' and `[m]any of the phenomena are indeed novel'.

Thus, dualities here point to the existence of new theories, but are ultimately bound to be superseded: the new theory, once found, will explain these dualities as being the result of certain approximations, which can be done in different ways, but lead to identical results, as articulated in the duality. But once that new theory is reached, the duality is no longer needed, except for practical purposes: for the resulting theory is a {\it single}, complete theory. In other words, establishing duality is here not the goal: rather, it is an intermediate step towards finding a new theory.

In what follows, I will dub that new theory, the one that supersedes the dual models and of which they are particular limits, the {\it successor theory}, $T_{\tn S}$.\footnote{The fact that there is a theory, $T_{\tn S}$, which succeeds the dual models, does not imply any specific stance about questions about scientific realism or referential stability across theory change. See footnote \ref{thloss}. For a summary of the debate over referential stability, see Radder (2012:~\S3.6) and Psillos (1999).} \\

These two viewpoints thus lead to different uses of duality in string theory. On the view discussed in Section \ref{exacteq}, the goal is to look for a theory, $T$, that realises the dualities as manifestly as possible. On the view in this Section, the goal is to find the successor theory, $T_{\tn S}$, that is ``behind'' the dualities, and which reveals them to be approximations. As I will argue in more detail in the next Section, even if they lead to two different research programmes, the two ideas need not contradict one another, and one could pursue both. 
%: it is well possible that both $T$ and $T_{\tn S}$ exist (at least in their respective regimes of validity), and that $T$ is an approximation to $T_{\tn S}$.  
But it is important to clearly distinguish the two functions: for otherwise, confusion easily ensues about the nature of duality, and about what one is entitled to expect from a duality conjecture.

%A case in point is AdS/CFT, the conjectured duality between gravity in asymptotically anti-de Sitter space (`AdS'), and a specific quantum field theory at the boundary of this space (CFT stands for `conformal field theory'). The traditional view, expressed in the quote by Aharony et al.~(1999) given in \S\ref{exacteq}, is that the string theory and the quantum field theory are exactly dual. This has also been expressed by Horowitz (2005:~p.~5), who goes further and sees the quantum field theory as one possible {\it definition} of non-perturbative string theory: `since the gauge theory is defined nonperturbatively [in AdS/CFT], one can view this as a nonperturbative and (mostly) background independent definition of string theory.'

%But other physicists think that this same duality, as given, is inexact. One of the arguments\footnote{J.~Maldacena and R.~Dijkgraaf, personal conversation.} is that the dual quantum field theory cannot account for certain entangled states of black holes in AdS, for which one may need more than a single CFT. This means that, at present, there is no consensus about whether AdS/CFT duality is supposed to be exact, as in the theoretical function (even though most of the official versions say that it is exact), or whether it is only an approximation to a successor theory, which is needed in order to properly define AdS/CFT. But these two views on the duality should be clearly distinguished, for they yield incompatible predictions for what one may expect from the duality.

\subsection{Does the distinction imply a tension?}\label{incompatib}

In this Section, I argue that the distinction between the two functions does not necessarily imply a tension. %(but not an incompatibility or contradiction {\it tout court}, of course!), on the simplest construal of the functions of duality, if expressed in sufficiently precise language.

At first sight, the previous quotes might suggest the distinction as a tension: in the first case (Section \ref{exacteq}), string theory and M theory instantiate the dualities exactly, while in the second case (Section \ref{dualheur}) dualities are perturbative clues towards finding a new theory, which will not instantiate duality exactly. However, one should interpret these quotations with some care, since they are not very precise (for example, the articles do not even include definitions of what is meant by `duality') and they involve quantum field theories and string theories which are still being developed: therefore, some of the central questions, viz.~whether the models as formulated are exactly valid, or whether dualities are exactly instantiated by the models, simply cannot be answered at this stage.

Nevertheless, I argue that the tension does not simply come down to lack of knowledge about the models involved: for the same tension exists for dualities and models which are exact, and well-known.\footnote{This tension is analogous to that between {\it emergence} and duality (cf.~De Haro (2015) and Dieks et al.~(2014)). In that case, there exist two mechanisms which can make duality and emergence compatible. First, there can be emergence {\it independently}, on the two sides of the duality (rather than across the duality relation). Second, there can be emergence if the duality is either not exact, or broken, for whatever reason. My resolution, in Section \ref{compare}, of the tension between the two functions of duality will be similar.}

Here are two important reasons why the two accounts, duality as exact equivalence, and duality as an approximately instantiated equivalence and pointing to new physics, might be thought to be in tension. First, they do not refer to two different levels of explanation or of ontology. Namely, being `two dual models of a single theory' or being `approximate dual models of a new underlying theory' both operate at the level of the formal structure: therefore, this potential resolution (`the two accounts operate at different levels, and so they do not contradict one another') is not available. Second, they might be seen to be in tension because the former sense assumes an exact duality, and being an exact instantiation of a theory; while the latter necessitates dualities which are not exactly instantiated, thus pointing to a new (unifying) theory, of which the two models are only approximations. 

Nevertheless, I claim that, when made explicit in a language sufficiently precise using the Schema from Section \ref{schema}, the tension turns out to be only apparent, and can be resolved. Namely, one distinguishes two {\it different theories}, corresponding to two different ways in which the theory to be constructed can relate to the given duality. Duality is then recognised as having two different {\it functions}, which aim at the construction of different kinds of theories, as I will analyse in \S\ref{compare}.\footnote{Coffey has noted an analogous tension in the context of theoretical equivalence between classical physical theories: a `tension between the symmetrical nature of theoretical equivalence and the asymmetrical nature of some reformulation judgments' (Coffey (2016:~footnote 19)). Though my tension differs from his (his focuses on asymmetry, and especially asymmetry of ontology; while mine focuses on theory construction), his questions are similar to mine: `One, how can a symmetric relation of theoretical equivalence accommodate an asymmetry of reformulation? Two, why does this asymmetry only occur in some cases and not others?' (p.~832). But our resolutions are very different: Coffey opts for an interpretative solution; but as I expounded in De Haro (2019), I think this is a mistake, since the tension is there even for uninterpreted theories.\label{coffeyh}}

\section{The Heuristic Function of Duality and Theoretical Equivalence}\label{compare}

In this Section, I come to the central question, of how the Schema of De Haro (2016, 2019a) and De Haro and Butterfield (2017), reviewed in Section \ref{schema}, bears on the heuristic function of duality and theoretical equivalence. I will illustrate, in some simple but explicit examples, how dualities (and symmetries) can be used heuristically. In Section  \ref{useh}, I will give examples of heuristic uses of approximate dualities, in the construction of new theories. In Section \ref{fheur}, I will analyse the resulting successor theories and models in more detail.

\subsection{How to use dualities heuristically}\label{useh}

In this Section, I will give examples of the use of dualities according to the heuristic function, i.e.~for constructing new theories. In \S\ref{tpp}, I will give an example of a point particle, and in \S\ref{aqg} I will make some analogies with similar problems in quantum gravity.

\subsubsection{Point particle heuristics}\label{tpp}

In this Section, I use the example of a point-like particle to illustrate the heuristic function. 
Our question is whether the successor theory, $T_{\tn S}$, which one constructs from a duality between models could be `bigger' than its models: what we would like the outcome of such a construction to be is a more general and precise theory, which comprises the models as {\it special cases}, or as {\it approximations} to, specific physical situations---so, they are approximate representations of the theory. 

It is not hard to suggest how this may happen, and I will illustrate this in one of the examples from De Haro (2019).\\

The example concerns a classical point particle moving on the real line. Its configuration space is ${\cal C}=\mathbb{R}$, and its space of states ${\cal S}$ is the cotangent bundle comprising the (canonical) position and momenta, i.e.~${\cal S}=T^*{\cal C}$. 

A model of this particle is given by a choice of polarisation of the cotangent bundle, i.e.~a local decomposition between (canonical) position and momentum variables, $X=(q,p)\in T^*\mathbb{R}$. Such a model is, as in Eq.~\eq{MnM}, a quadruple:
\bea
M_X:=\bra T^*\mathbb{R}_X,\o_X;H_X;E_X;X\ket~.
\eea
The specific structure is here the local decomposition, $X$, between position and momentum variables. The state space is ${\cal S}=T^*\mathbb{R}_X$, equipped with a symplectic form. The set of quantities, ${\cal Q}$, contains a single element, namely the Hamiltonian, $H_X$ (for simplicity of the model, I have taken this to be the only quantity: but it is of course no problem to include any powers of $x$ and $p$ also as quantities in ${\cal Q}$). $E_X$ is the dynamics, presented as the Hamiltonian equation of motion. 

But any other choice of such a decomposition, $\bar X=(\bar q,\bar p)$, preserving the Poisson bracket $\{X^\a,X^\b\}=\o^{\a\b}$ (where $\a,\b=1,2$ and $X^1=x$, $X^2=p$), will of course give an equivalent model. In other words, a change of polarisation, given as a linear map $S:X\mapsto\bar X=S\cdot X$, gives an equivalent model iff it preserves the symplectic (closed and degenerate) two-form $\o:=\dd q\wedge\dd p$. The set of transformations $S$ satisfying these conditions turn out to act on $T^*\mathbb{R}$ as $\mbox{SL}(2,\mathbb{R})\cong\mbox{Sp}(2,\mathbb{R})$, viz.~the group of area-preserving linear transformations on the phase space. The models are therefore best presented, alternatively: as given by the action of an element $S\in\mbox{Sp}(2,\mathbb{R})$ on some fidutial, i.e.~initially chosen, polarisation; as follows:  
\bea\label{Ms}
M_S:=\bra T^*\mathbb{R}_{\bar X},\o_{\bar X};H_{\bar X};E_{\bar X};S,X\ket_{S\in{\sm Sp}(2,\mathbb{R})}~,
\eea
where we now have one model for each $S\in\mbox{SL}(2,\mathbb{R})\cong\mbox{Sp}(2,\mathbb{R})$.
Here, $\bar X=S\cdot X$, as above, and the subscript ${\bar X}$ indicates that the corresponding item is evaluated using $\bar X:=S\cdot X$. The specific structure, $\bar M=\{S,X\}$, is given by the particular $S\in\mbox{Sp}(2,\mathbb{R})$ matrix chosen for the model, together with the choice of reference state $X$, out of which the set of models is generated. 

In De Haro (2019) it was argued that one can reconstruct a common core theory from a set of models Eq.~\eq{Ms}. The theory is obtained by taking the union of all these models, and modding out by an equivalence relation which identifies two models if they belong to the same $\mbox{Sp}(2,\mathbb{R})$ orbit, i.e.~if there is an $S\in\mbox{Sp}(2,\mathbb{R})$ which relates the variables $\bar X$ between the two models. The result is of course a set of states which is the symplectic manifold ${\cal S}=(T^*\mathbb{R},\o)$, together with a single scalar quantity, and the Hamilton equations, written in terms of $\o$ (see e.g.~Abraham and Marsden (1978:~Chapter 3), Butterfield (2006:~\S4.3)).\\

%But the theory thus obtained is, like in the two examples discussed at the beginning of this Section, not the kind of theory we are after with the heuristic function of dualities. For it is simply a theory of which each of the models $m_S$ is an exact representation, and all the models are dual to each other in an exact way. 

%One should also notice that, in this example, the duality ends up being just a symmetry of the model (which is a trivial case of duality, i.e.~self-duality of the model). This is because in the final formulation there is a single state space, a single set of quantities, and a single dynamics: even at the level of the models. The state spaces, quantities, and dynamics in each of the models are representations of this single quantity, but in a very trivial way: they are not what one would call `equivalent representations', but rather they are {\it rewritings} of the {\it same} representations.\\

In geometric quantisation, the $\mbox{Sp}(2,\mathbb{R})$ invariance of the classical theory carries over to the quantum theory, where the Poisson bracket is replaced by a commutator, $[X^\a,X^\b]=i\hbar\,\o^{\a\b}$. A {\it state} is now specified by an element of the Hilbert space of square-integrable wave-functions on the joint spectrum of a maximal set of commuting operators. Picking a basis for this Hilbert space is choosing a polarisation of the wave-function, which is allowed to depend on both $x$ and $p$, $\psi(x,p)$, but is subject to an additional constraint. For example, demanding ${\pa\psi(x,p)\over\pa p}=0$, we get the position representation of the wave-function. 

This choice is of course a stipulation, and it is not unique: any other choice of polarisation can be made by picking an element $S\in\mbox{SL}(2,\mathbb{R})$, and so $\bar X:=S\cdot X$, such that $X$ and $\bar X$ are linearly independent, and demanding the independence of the wave-function upon this particular element: 
\bea\label{polar}
{\pa\psi(X,\bar X)\over\pa \bar X}=0~.
\eea
This requirement thus builds the $\mbox{Sp}(2,\mathbb{R})$ duality (which includes the $x\leftrightarrow p$ duality) into the common core theory.\\

The above use of $\mbox{Sp}(2,\mathbb{R})$ duality suggests how to get a successor theory describing more degrees of freedom, of which the above models, $M_S$, are only {\it approximately} models, i.e.~representations. Each of the models should somehow have an embedding in the successor theory (and perhaps even an {\it extension} into that theory), so that duality does {\it not} hold exactly in the entire new theory, and it does {\it not} hold exactly between the extensions of the models, if such extensions are given. The idea is indeed to break the duality.

We can obtain such a theory by a slight modification of the quantum mechanical point-particle example. The physics we will be entertaining here is slightly speculative: but notice that this is exactly what the heuristic function of dualities is supposed to do!---it is supposed to help us construct new theories. So, by suggesting, in this Section, how the simple point-particle example can lead to new theories, I hope to illustrate the heuristic function of dualities in the string theory and M theory programme: which is, of course, immensely more complex.

The idea is to study a specific quantum version of the classical particle: not just a straightforward quantisation of a point particle, but a quantum theory which includes additional dynamics.

The heuristic approach to this duality suggests a generalisation of this independence of the wave-function from half of the variables. After all, that constraint arises from the kinematically given algebra between position and momenta. But the dynamics might dictate a more general algebra, of which the simple Heisenberg algebra is a special case. In other words, the heuristic approach here seeks to make the choice of polarisation arise dynamically from the theory, and perhaps receive corrections away from some `perturbative' limit. So, the right-hand side of the Eq.~\eq{polar} only goes to zero in a special limit. The Hilbert space then has two sectors (position and momentum), one of which is dropped by the choice of polarisation (which now arises as an equation of motion, in a special limit). But in the full theory, without taking any limits, the entire Hilbert space is required. The $\mbox{SL}(2,\mathbb{R})$ symmetry is thus a special simplifying property of the limit. More precisely: we are led towards a theory with a Hilbert space that is constructed from $L^2(T^*\mathbb{R})$ rather than $L^2(\mathbb{R})$. The latter will arise as a limit of the theory, in which one of the equations imposes a choice of polarisation. In the next subsection I give some examples, from the quantum gravity literature, where such things happen.

\subsubsection{Quantum gravity heuristics}\label{aqg}

In this Section, I discuss analogues, from the quantum gravity literature, of the idea discussed in the previous Section, namely of doing a geometric quantisation of a point particle which is subject to a duality constraint like Eq.~\eq{polar}, and then generalising this by including dynamical corrections that break that duality.

%In some cases, $Z(X)$ and its complex conjugate\footnote{In some cases, the bar indicates a certain property of the representations of the gauge group involved, called `chirality'. See AOSV (2004:~Eq.~(5.4)).} $\bar Z(\bar X)$ describe, together, the perturbative sector of a larger theory (e.g.~a theory of black holes), constructed from the two theories: $Z_{\tn{BH}}=|Z(X)|^2$. According to one interpretation (Kraus (2006:~\S2)), the two  sectors correspond to the chiral and anti-chiral sector of the theory, which are completely decoupled. 

%However, in some cases, e.g.~in the presence of a so-called `gravitational anomaly', the partition function gets quantum corrections that require additional summation over moduli: $Z_{\tn{BH}}=\sum_\a|Z_\a(X)|^2$. For example, this can happen in the presence of a gravitational anomaly of the black hole theory. In such a case, the central charges of the chiral and anti-chiral field theories are not equal (the difference in central charges being the gravitational anomaly, Kraus (2006:~Eq.~(3.23)).

A first example closely following the previous discussion  are quantum theories for the point particle which incorporate quantum gravity effects. Here, the mechanism for breaking the $\mbox{SL}(2,\mathbb{R})$ duality is different. The classical theory is quantised by demanding $[X^\a,X^\b]=i\hbar\,\o^{\a\b}$ like before, but quantum gravity effects, such as gravitational particle collisions, correct this relation, adding $X$-dependent terms on the right-hand side.\footnote{For a seminal discussion of such commutators, see Snyder (1947:~Eq.~(8)). For more recent accounts in the context of quantum gravity, see Kempf (1993:~\S2.1), De Haro (1998:~Eq.~(69)). For generalised versions of the uncertainty principle, see Maggiore (1993:~Eq.~(7)).} In this case, too, the $\mbox{SL}(2,\mathbb{R})$ duality is explained as an approximate duality of the successor theory, which becomes exact in the limit in which quantum gravity effects are negligible; but away from that limit, the theory looks quite different, and the wave-function does not satisfy the simple constraint Eq.~\eq{polar}. The upshot is that duality can here be understood as the result of a `turning off' of quantum gravity effects. 

%The second analogy is with the phenomenon of the `holomorphic anomaly', in string theory. The analogue entails two complex variables $X$ and $\bar X$ which are each other's complex conjugates. The constraint Eq.~\eq{polar} is then a holomorphicity constraint on the wave-function $\psi(X,\bar X)$. Quantum mechanical effects can break the holomorphic constraint Eq.~\eq{polar}, getting a correction term on the right-hand side. The wave-function then depends on both $X$ and $\bar X$ (Bershadsky et al.~(1993:~\S3)).

{\it S duality.} The example of a point particle, and of duality as an $\mbox{SL}(2,\mathbb{R})$ symmetry, is also reminiscent of S duality in quantum field theory, which is a generalisation of electric-magnetic duality, in four-dimensional (supersymmetric) Yang-Mills theory. In that case, the duality group is $\mbox{SL}(2,\mathbb{Z})$, and it acts as integral-fractional transformations on the theory's complexified coupling, which takes values on a two-torus, $\mathbb{T}^2$. Notice that, despite the fact that this might look similar to a gauge symmetry, it is not a gauge symmetry at all, because it is not the gauge fields, but the coupling constant which is being transformed. 

The duality group $\mbox{SL}(2,\mathbb{Z})$ can be given a geometric meaning, by embedding the theory in six dimensions, i.e.~considering the manifold which is the product of the two-torus (the space on which the coupling takes values) with ordinary four-dimensional Minkowski space, i.e.~$\mathbb{T}^2\times\mathbb{R}^4$. The coupling constant can then be identified with the complex structure of the two-torus $\mathbb{T}^2$, and the duality group $\mbox{SL}(2,\mathbb{Z})$ becomes the mapping class group of, what is now, a real torus. 

The low-energy dynamics of this six-dimensional gauge theory reproduces that of the four-dimensional super-Yang-Mills theory, with its duality symmetry $\mbox{SL}(2,\mathbb{Z})$. But now once the six-dimensional interpretation is reached, one can imagine more general situations (of high energies, and-or containing other kinds of interactions) in which the four-dimensional perspective is only an {\it approximation} to the six-dimensional dynamics, so that also the four-dimensional duality group $\mbox{SL}(2,\mathbb{Z})$ is only an approximate symmetry. 

Thus, here again, making the duality group arise as an approximation of a physical system, suggests generalisations to a successor theory---in this case, a higher-dimensional theory---which contains more possibilities than the ones strictly postulated by duality. 

\subsection{Successor theories and models}\label{fheur}

In this Section, I gather the ideas from the examples in \S\ref{useh}, and give a more concrete account of the successor theory, $T_{\tn S}$, according to the schema for duality introduced in Section \ref{schema}. I first discuss the point particle (\S\ref{pp2}) and then discuss successor theories more generally (\S\ref{successorth}). 

\subsubsection{Theories and models: the point particle}\label{pp2}

We briefly go back, in this Section, to the point particle example from Section \ref{tpp}, and its quantum gravity analogue from Section \ref{aqg}, to discuss the resulting successor theory. The idea was to consider initially a theory with a simple duality, namely the group $\mbox{SL}(2,\mathbb{R})$ or $\mbox{SL}(2,\mathbb{Z})$, expressing a redundancy of the theoretical description, so that some putative degrees of freedom are unphysical and so are constrained by the duality. On this common core theory, duality is realised as a symmetry, and the difference between taking a wave-function that depends on a coordinate $X$, and taking a wave-function that depends on a different coordinate $\bar X:=S\cdot X$, where $S\in \mbox{SL}(2,\mathbb{R})$ (or $\mbox{SL}(2,\mathbb{Z})$) is a symmetry transformation, was a mere `choice of polarisation', by the theory's own lights. However, further physics could lead one to a different situation, described by the successor theory, in which the duality symmetry is broken. Thus, physical effects came to distinguish two choices, $X$ and $\bar X$, as physically distinct situations after all: so that the theory therefore changes, because the symmetry constraint, Eq.~\eq{polar}, is no longer satisfied---it is replaced by a more general formula in which the symmetry is allowed to be broken. For example, in the case of S duality, the $\mbox{SL}(2,\mathbb{Z})$ transformations are the symmetries of a two-torus (they correspond to distinct complex structures of the torus), and they correspond to geometrically distinct six-dimensional manifolds (cf.~\S\ref{aqg}). Manifolds with distinct complex structure are indistinguishable at low-energies, but at high energies the different tori can be distinguished.

Because the initial models were consistent, there are well-defined physical conditions under which $\mbox{SL}(2,\mathbb{R})$ is indeed a duality group of the common core theory---thus realising duality just as  symmetry, which entitled one to interpret the models as representing physically identical situations (under certain assumptions which we do not need to consider here).\footnote{For a discussion of symmetries, cf.~Greaves and Wallace (2013), Teh (2016), De Haro et al.~(2015). It is fair to say that there is no general {\it formal} criterion known that could a priori tell us whether a symmetry contains physical content. (For a review of several failed attempts at finding a formal notion of equivalence that relates physically identical states of affairs, see Belot (2013). For a view on symmetries where the formal and the interpretative aspects are in balance, see Caulton (2015).) Judgment about this surely requires looking at the specific details of the theory in question, and at various messy physical details of the system to which the theory is applied.\label{symmy}} But without these physical conditions, i.e.~if the $\mbox{SL}(2,\mathbb{R})$ symmetry is broken by dynamical effects as just discussed for the torus, then an  $\mbox{SL}(2,\mathbb{R})$ transformation relates physically distinct situations. The theory which distinguishes the $\mbox{SL}(2,\mathbb{R})$-distinct situations is the successor theory, $T_{\tn S}$ (i.e.~the theory which includes the corrections on the right-hand side of Eq.~\eq{polar}). When duality is realised in this theory as an approximate symmetry, the symmetry transformation can be interpreted physically: not as a redundancy,\footnote{Recall, from footnote \ref{symmy}, that not all symmetries express redundancies. But, for simplicity of exposition, I assumed that the symmetries considered in Section \ref{tpp} express sameness of description, i.e.~the corresponding interpretations are internal, in the sense of De Haro (2016).} but as making an actual physical distinction! And so, we have here a miniature version of the kind of role duality is supposed to play in the string and M theory programme.

Notice that the successor theory, $T_{\tn S}$, also gives an {\it explanation} for the duality: namely, in the duality, what are in fact {\it different models}, or different limits of $T_{\tn S}$, look like isomorphic models $M_i$, because of the approximations (to $T_{\tn S}$) they introduce.

The model notation $M_i=\bra m_i,\bar M_i\ket$ (for $i\in I$ in an appropriate index set $I$) from Eq.~\eq{MnM} should make this clear. For example, in the point particle case, equivalent models were obtained by $\mbox{SL}(2,\mathbb{R})$ transformations, and so the set indexing the different models is $I=\mbox{SL}(2,\mathbb{R})$. The specific structure was given in that case by, first, the specific $\mbox{SL}(2,\mathbb{R})$ transformation used and, second, the representative, $X$, of the orbit, so that: $\bar M=\{S,X\}$, and $S\in\mbox{SL}(2,\mathbb{R})$. In this case, all the models are isomorphic to each other, i.e.~$m_i\cong m_j$ ($i,j\in I$). But this is in general of course not so, because a theory can have both equivalent and inequivalent models. The set of models $\{M_i\}_{i\in I}$ (isomorphic and non-isomorphic ones) is then the set of representations of the underlying theory, $T$. 

The heuristic construction of the successor theory, $T_{\tn S}$, is then envisaged to proceed in two steps, as follows:\footnote{There is no claim here that the two-step procedure outlined below is somehow unique or necessary. One can think of slightly different procedures that amount to the same thing: namely, a successor theory $T_{\tn S}$ which is approximated by both $T$ and by $T$'s models.}

Initially, i.e.~for exact dualities, and assuming for simplicity only equivalent models, only the model triples $m_i$ ($i\in I$), i.e.~the models stripped of their specific structure, are interpreted as being physical. More precisely: the specific structure, $\bar M_i$, gives each model its specificity: it is like the choice of $\mbox{SL}(2,\mathbb{R})$-representative $S$, and it is not physical. Because Eq.~\eq{polar} holds, such a choice does not correspond to actual physics, but is made by stipulation.

But subsequently, some new physics modifies the relation like Eq.~\eq{polar} (i.e.~the relation which allowed us to stipulate a choice of specific structure), so that (typically!) more variables are needed to describe the problem: which prompts us to interpret part of the specific structure as actually being physical, in the modified model. In the example of S duality, the set $I=\mbox{SL}(2,\mathbb{Z})$ now receives a physical interpretation as the mapping class group of a torus,\footnote{The mapping class group of a topological space is the discrete group of symmetries of that space.} $\mathbb{T}^2$, which is part of a physical, six-dimensional geometry. This choice is now physical, because the choice of complex structure of $\mathbb{T}^2$ is physical---it is part of the geometry on which the theory is defined. And so, a choice of specific structure---a choice of $S$---is no longer innocuous; and Eq.~\eq{polar} receives corrections, which correspond to different such choices. The successor theory, $T_{\tn S}$, describing these corrections differs from $T$, because it incorporates the symmetry as a {\it physical} symmetry of its triple, at least in a suitable approximation (such as: the area of the internal torus going to zero), which now incorporates some of what used to be the specific structure. In effect, we have changed the models and the theories: some of the specific structure has now become part of the triple, giving rise to new states (new physical situations), new quantities (which make a distinction between those situations) and new dynamics (accounting for new interactions). 

\subsubsection{Successor theories and the heuristic function}\label{successorth}

The heuristic function of dualities is the ability to use dualities in the constructive way just discussed, i.e.~for building new theories: starting with exact dualities, viz.~equivalent models, building successor theories that implement the duality as approximate symmetries (or as other constitutive parts of the theory's structure leading to approximately isomorphic models), and then reinterpreting the symmetries, or parts of structure, as special properties of a limit or approximation to a specific physical system. Away from the limit, the number of degrees of freedom of the theory (i.e.~number of the states and-or quantities) is typically not reduced, but increases: at any rate, the physical interpretation changes. 

Thus there is no longer a duality, but only a theory with one or several approximations or limits: duality holds only approximately, but there is a self-consistent regime in which duality obtains. The use of $T_{\tn S}$, from the point of view of duality, is that it explains the physical origin of the duality, and exhibits how duality is implemented. So, duality ends up being a property of idealised models, but not a property of the physical successor theory. This is what Radder (1991) has called heuristics `from the old theory to the new theory', i.e.~the heuristic function helps one find a successor, more accurate, theory given a set of models.

The conceptual picture arising should now be clear, in the Schema from Section \ref{schema}. We have an initial theory, $T$, and its set of isomorphic models, $M_i$ ($i\in I$). The theory and its models need not be well-defined for arbitrary values of the parameters; they may have limited validity, and also the dualities may hold only approximately  (in an idealisation within the successor theory, in which certain interactions, or certain complicating factors such as finite area, are neglected). The successor theory, $T_{\tn S}$, is then able to reproduce $T$ and its models (or something very close to them) as special cases, for particular approximations. $T_{\tn S}$ does not exhibit exact duality, and the models are not exact representations of $T_{\tn S}$. Also, $T_{\tn S}$ usually reinterprets the specific structure $\bar M_i$ of the models physically: $T_{\tn S}$ then {\it changes the definition of the models}.\\

%Going back to the similarity between the tension of the theoretical vs.~the heurstic functions, with that between emergence and duality, mentioned in (1) of Section \ref{tat}, 
We see that there are two ways to make the theoretical and the heuristic functions compatible, i.e.~we can:\footnote{This is analogous to the resolution of the tension between duality and emergence, in De Haro (2015) and Dieks et al.~(2014).}

(i) Extend the models beyond their original domain of application, even if they are no longer exactly dual, and find a successor theory, $T_{\tn S}$, of which those models, perhaps modified, are now exact (but not necessarily dual) representations; or 

(ii) We can simply find a new theory, $T_{\tn S}$, of which the original models are approximate representations. 

Intermediate positions are of course also possible. %In both cases, this amounts to the original duality's being {\it not exact} (i.e.~there is equivalence only within a specific range of parameters, or for specific physical situations or systems, or for particular mathematical approximations) and {\it invalid} in the successor theory $T_{\tn S}$, i.e.~not instantiated by it (because only approximately valid there).

The analogy with the analogous problem of symmetries, mentioned in footnote \ref{symmy}, is that the physical contents of $T$ and $T_{\tn S}$ are different. In $T$, one was entitled (though not invariably obliged) to interpret dualities as mere redundancies. This is no longer possible in $T_{\tn S}$, because the duality is now seen to be a consequence of an approximation to a certain physical situation: in other words, the physical contents of $T$ and $T_{\tn S}$ are different, and part of the physical content of $T_{\tn S}$ now implies the approximate duality.

\section{Comparing the Schema with Other Philosophical Work on Duality}\label{compareo}

In this Section, I will compare the Schema's analysis of the heuristics of duality, to other accounts in the recent philosophical (and physical) literature. In Section \ref{igsrev}, I compare the present analysis to another interesting account of duality, namely the `duality-as-gauge-symmetry' account. In Section \ref{ricklesd}, I compare my construal of the successor theory, $T_{\tn S}$, with what Rickles has called the `deeper theoretical structure' that is behind a duality.

\subsection{Isomorphism vs.~gauge symmetry and heuristics}\label{igsrev}

In this Subsection, I let the analysis of the heuristic function from Sections \ref{dphysics} and \ref{compare}, which was an application of the Schema from Section \ref{schema}, bear on the comparison between the isomorphism vs.~the gauge symmetry accounts of duality,  for the heuristic function.\footnote{The gauge symmetry account of duality is also discussed in De Haro (2019a), for the theoretical function.}

If the Schema's account of duality can be stated (in a slogan) as `duality is an isomorphism of model roots', then the gauge symmetry account can be stated (in a slogan) as `duality is a gauge symmetry of a deeper theory'.

According to the gauge symmetry account, a duality points toward the existence of a theory which realizes the duality as a gauge symmetry. Now `gauge symmetry' is a vague word about which there is still much confusion. It is therefore necessary to make the following distinction (quoting from De Haro et al.~(2015:~Section 2, (i)-(ii))):\footnote{`Gauge symmetry' does not at all exhaust the meaning of `symmetry'. My argument is that, although in most cases duality can be seen as a symmetry of $T$, it is in general not a {\it gauge} symmetry. }

`(i) (Redundant): If a physical theory's formulation is redundant (i.e.~roughly: it uses more variables than the number of degrees of freedom of the system being described), one can often think of this in terms of an equivalence relation, `physical equivalence', on its states; so that gauge-invariant quantities are constant on an equivalence class and gauge-symmetries are maps leaving each class (called a `gauge-orbit') invariant. Leibniz's criticism of Newtonian mechanics provides a putative example: he believed that shifting the entire material contents of the universe by one meter must be regarded as changing only its description, and not its physical state.

(ii) (Local): If a physical theory has a symmetry (i.e.~roughly, a transformation of its variables that preserves its Lagrangian) that transforms some variables in a way dependent on spacetime position (and is thus `local') then this symmetry is called `gauge'. In the context of Yang-Mills theory, these variables are `internal', whereas in the context of General Relativity, they are spacetime variables.'

The trouble with the (i) (Redundant) account of duality (i.e.~the idea that duality is a gauge symmetry, where gauge symmetry is construed as (i) (Redundant)), is that it wrongly assumes dual theories to be invariably physically equivalent. I have argued this point in detail in De Haro (2016:~\S1.3, \S2.2) and I will not dwell on it here. Namely, concluding that two dual theories are invariably physically equivalent without further technical and philosophical analysis, is simply incorrect: as one can show with obvious counter-examples (and this is true whether one construes the phrase `physical equivalence' in terms of reference, as I do, or in terms of the descriptive capacities of a theory).\footnote{The gauge symmetry account, especially in its (i)~(Redundant) variant, is defended by Rickles (2017): `this can be generalized to {\it all} cases in which one has a duality symmetry: they can always be promoted to gauge-type symmetries because they just {\it are} gauge-type symmetries' (p.~66). In terms of the distinction just given, (i) vs.~(ii), Rickles' account is, however, ambiguous. Overall, he seems to endorse the more general notion of gauge symmetry, viz.~(i)~(Redundant). For example, he mentions that a `natural interpretation [of duality] should be taken as a mere representation of a deeper underlying structure' (p.~63), and so the emphasis present in `mere representation' suggests he means (Redundant). Also, he clearly characterises his proposal as follows: `the differences [should] be viewed as unphysical in exactly the sense of gauge freedom: the duality mapping leaves invariant all of the physically meaningful quantities and symmetries' (p.~65). However, elsewhere he mentions, as an example of a gauge symmetry, the ability to choose a specific labelling of points in general relativity (p.~64): which would rather make it a case of (Local).  Also, when he quotes Fuchs and Schweigert on p.~66, he characteries their quotation as being about `gauge symmetry', whereas the original passage quoted refers to a more general kind of symmetry, not a gauge symmetry. Overall, however, Rickles seems to have (i)~(Redundant) in mind.}\\

The construal of gauge symmetry I am therefore interested in here is (ii) (Local). First: because this is an interesting proposal in its own right, and some examples in string theory would seem to point in that direction.\footnote{The case of T duality can, on a sufficiently technical and perspicacious formulation, indeed be regarded as (part of) a gauge symmetry. More precisely, the discrete $\mathbb{Z}_2$ duality group that is T duality, can be shown to be a subgroup of an $\mbox{SU}(2)\times\mbox{SU}(2)$ gauge symmetry formed by the operators describing the bosonic background of the theory (cf.~Polchinski (1998:~pp.~242-248)). The question thus naturally arises whether this generalises to all dualities (to which my answer, from the $\mbox{SL}(2,\mathbb{R})$ and $\mbox{SL}(2,\mathbb{Z})$ examples, and further analysis below, is: No).} Second, it is an interesting alternative to the Schema, especially because it is technically much more specific than (Redundant): its level of specificity roughly matching the most detailed version of the Schema in Section \ref{schema}, i.e.~duality as an isomorphism of triples of states, quantities, and dynamics, which are representations of the same theory, in the representation-theoretic sense of the word.

The example of the point particle in Section \ref{tpp} might give the {\it false} impression that the gauge symmetry account, (ii) (Local), is indeed confirmed, because duality ends up being an $\mbox{SL}(2,\mathbb{R})$ or $\mbox{SL}(2,\mathbb{Z})$ symmetry of the theory. But, as already remarked in Section \ref{aqg} for the case of S duality: these are not {\it gauge} symmetries (in the sense of (Local)) at all! %Also in the example of Section \ref{nonheur}, which dealt with the theoretical function of dualities only, there was no gauge symmetry. 
On the other hand, these both examples {\it are} cases of isomorphism between models, i.e.~they confirm the Schema in Section \ref{schema}.

In fact, there is a stronger conclusion that follows from this. The summary (i)-(ii), in the previous Section, of the two ways of realising the heuristic function, does not make any reference to a theory $T$ instantiating the duality as a symmetry (whether gauge or otherwise). Indeed, although an instantiation of duality as a symmetry of $T$ may exist some cases, it is not the general case: namely, if $T$ exists, it need not realise duality as symmetry. But neither is duality realised as a symmetry of the successor theory, $T_{\tn S}$: in fact, we know that duality {\it cannot} be a symmetry of $T_{\tn S}$! For if it were, the duality would be exactly instantiated by $T_{\tn S}$---which by definition it is not, either on the first account (i), or on the second, (ii), from the previous Section. In other words, {\it there is no requirement that $T$ or $T_{\tn S}$ must realise duality as symmetry} (gauge or otherwise): and, in fact, it is also true that {\it duality is not in general realised as a symmetry of $T_{\tn S}$,} but only (if at all) as a symmetry of $T$ (though not a {\it gauge} symmetry). Thus the `duality as gauge symmetry' account, though at first sight appealing, is unable to deal with these examples.

\subsection{Rickles' comments on the `deeper theory'}\label{ricklesd}

I mentioned, in footnote \ref{except} of the Introduction, that Rickles (together with Dieks et al.~(2014) and De Haro and Butterfield (2017)) is the exception, in the recent literature on dualities, to the statement that the heuristic use of duality has gone unnoticed. In this Section, I compare the Schema's analysis of the heuristic function to a few of Rickles' comments.\footnote{I thank one reviewer for providing me with the set of quotes below.} Rickles' ideas here, as on many other issues, are deep. But I will also find them to be problematic.

I will comment, below, on a number of passages by Rickles (all emphases are mine), which revolve around two main points, both of which are supposed to highlight the heuristic function of duality:\\

(i)~~\,Dualities point towards a `deeper theoretical structure' that explains why there is a duality.

(ii)~~A good analogy for this deeper theoretical structure is given by the electromagnetic field, which unifies the electric and magnetic fields.\\

I will of course agree with Rickles on his first point, (i): since this is what I have advocated in detail in this paper, and indeed also briefly in previous papers (see e.g.~Figures 2-4 in Dieks et al.~(2014)). Thus roughly, Rickles' phrase `deeper theoretical structure' could tentatively be taken to correspond to my successor theory, $T_{\tn S}$, which it is the task of the heuristic function to discover (as opposed to the theoretical function, which discovers the theory $T$). 

But I will argue that (ii) is mistaken: for the electromagnetic field is not an example of the use of the heuristic function at all. Rather, it is an example of the theoretical function (the analogy between gauge-gravity duality and electric-magnetic duality is discussed in detail in Section 3.2.2 of Dieks et al.~(2014), as part of the theoretical function). Thus I will argue that Rickles' `deeper theoretical structure' sometimes refers to $T_{\tn S}$, and sometimes to $T$: and so, his phrase `deeper theoretical structure' does not really express the heuristic function of duality.\\
 
My first four quotes from Rickles below will add up to point (i), i.e.~that there is a deeper theoretical structure that explains why there is a duality:\\

(1)~~`The mere existence of a duality points to unphysical structure in the dual theories. Thus, any interpretation we give is `provisional', and points towards some common core. {\it This common core might be fully understood (as a deeper theoretical structure encompassing both dual theories) or might be known only via the limited information provided by the dual pair.}' (Rickles (2017:~p.~66)).

(2)~~`Ultimately, of course, duality is not such a good case study for those who wish to deny fundamentalism, since {\it it points to deeper structures that may or may not themselves have fundamental status}' (Rickles (2011:~p.~66)).

(3)~~`The fact that the {\it dualities have been used to discover genuinely new and unexpected physics} are enough to pose a problem for anti-realists who will need to provide an explanation for how this is possible.' (Rickles (2011:~p.~66)).

(4)~~`[T]here is deep structure (given by the shared symmetries) that the two theories have in common. {\it This characterises a deeper structure, not yet fully understood, that will admit at least two representations.}' (Rickles (2013:~p.~319)).\\

The next two quotes, below, make the point, (ii), that, in the example of electric-magnetic duality, the electromagnetic field (which is the deeper theoretical structure in that case) integrates both the electric and the magnetic field, so that they are two aspects of one structure:\\

(5)~~`In this case {\it the duality points to a deeper structure into which both the electric and magnetic fields are integrated, namely the electromagnetic field}. Hence, the discovery of a duality between a pair of things can be `symptom' that the pair of things are really {\it two aspects of one and the same underlying structure}' (Rickles (2011:~p.~57)).

(6)~~`[S]uch identification of core structure provides a {\it methodology for scientific discovery: identify common structures between theories or structures and then try to understand this common structure via another deeper, broader theory that admits of multiple representation.} This is just what we find in the case of the {\it duality between electricity and magnetism which leads us to the deeper electromagnetic field.} (Rickles (2013:~p.~320)).\\

As I said above: quotes (1) to (4) are, at least in letter, along the lines of what I have said in this (and earlier) papers; and so I agree with Rickles. But my criticism is as follows:\\
\\
(I)~~{\it No philosophical analysis of the heuristic function.} That dualities point to deeper theoretical structure is of course known. For it was indeed the main claim of Witten's breakthrough lecture and paper in 1995 (see the quotes in Section \ref{dualheur}), namely the launching of the M theory programme.\footnote{Rickles does of course contextualise the physicists' claim within his other philosophical remarks. Also, Rickles' work on dualities has other virtues: especially, his connecting duality to some of the philosophical discussions on theoretical equivalence. Finally, one must credit Rickles for pointing out, in quotation (6), that there is a question of methodology in connexion with dualities.}\\
% What I {\it am} saying here is that Rickles' mention of the heuristic function adds nothing to what the physics literature has already repeatedly said.\\
\\
(II)~~{\it Confusion between the theoretical and the heuristic function.} Although Rickles repeatedly uses a phrase like `deeper theoretical structure' or `deeper structure', he does not say exactly what he means by it: and, in fact, in the few instances in which he does say it more explicitly, he gets it backwards. Namely, he refers to the theoretical rather than to the heuristic function. This is seen from the following four points, of which (A) and (D) are the central ones: \\

(A)~~In quote (4), he uses the word `representation'. But the definition of the successor theory, in Section \ref{dualheur}, means that the dual models are {\it not} representations of the successor theory $T_{\tn S}$! They are representations of $T$ (if this theory exists), and only {\it approximate} representations of $T_{\tn S}$. Thus it seems that Rickles has here something like the theoretical function of duality in mind, rather than the heuristic function.\footnote{This argument is independent of whether or not one endorses the Schema from Section \ref{schema}. For the statement that the dual models cannot be representations of $T_{\tn S}$ just follows from the notion of the successor theory.}

(B)~~Also, in the quoted passages, Rickles never uses words like `approximation', `inexact or inaccurate theory', `perturbative expansion', etc.: which are key words to describe what the heuristic function is about (see the quotes by the physicists in Section \ref{dualheur}). Had Rickles used them, he {\it would} have made it clear that he was talking about the heuristic function. 

(C)~~Rickles' phrase `deeper structure' is of course in itself ambiguous, because it could refer either to $T$ (where `deeper' is used as in `mathematically more perspicuous',  `more sophisticated', or `more amenable to generalisation'), or it can also refer to $T_{\tn S}$ (where `deeper' is used as in `valid in a physically larger domain', or `valid at higher energies'), i.e.~it can refer either to the theoretical or to the heuristic function, at least without further elaboration. 

(D)~~His example of electromagnetism is incorrect, because {\it the electromagnetic field describes exactly the same physics as do the electric and magnetic fields}.\footnote{This is of course no longer true in quantum field theory, where the gauge field formulation is more fundamental, like for non-abelian theories. But this is not the issue here, since Rickles is talking about classical theories.} Compared to the electric and magnetic fields, the electromagnetic field is not a case of a `deeper' theory, in the sense of an approximation like $T_{\tn S}$. Rather, it is deeper in the sense of the theoretical function leading to $T$, i.e.~in the sense of giving a more perspicacious formulation (because mathematically more sophisticated, more unified, because the symmetries are explicitly exhibited, etc.) of the same physics. 

Thus, (A) to (D) amount to lacking a clear idea about the distinction between the theoretical and the heuristic function. 

%What seems to me to be a worthwhile philosophical project for heuristics and methodology is rather: 

%(i)~~To analyse the heuristic function, in particular explaining how dualities achieve new physics. 

%(ii)~\,To do this against a philosophical background discussion about heuristics and scientific methodology and practice.

%This is the project I have attempted to undertake in this paper.

\section{Discussion and Conclusion}\label{conclusion}

In this paper, I began by making a distinction between a theoretical and a heuristic function of duality and of theoretical equivalence. Then, using the Schema from De Haro (2016, 2019a) and De Haro and Butterfield (2017), I described these functions in detail, as follows. The aim of the theoretical function is to discover, or to (re-)construct, theoretically equivalent or dual models (theories). Thus the result achieved using that function is a set of models, of which a number are theoretically equivalent (dual), together with the common core theory---a theory of which those models are exact representations. The theoretical function can be stated in precise mathematical terms, and in some cases there are even reconstruction theorems that allow one to reconstruct a theory from its set of models. The question here is not primarily about new physics: for a theory thus constructed describes, at least in principle, the same systems as do the models from which it was obtained. Nevertheless, a common core theory is formally more perspicuous, and can also have other advantages over the models from which it was constructed: for example, it admits new interpretations, or it admits new models. %In this sense, the theoretical function is a tool towards theory construction.

The heuristic function aims to discover a {\it successor theory}, in the innocuous sense of `a theory whose content goes beyond the content of the original models, and of which the dual models are only approximate instantiations'. The relation between this successor theory and its dual models, and the status of duality (or theoretical equivalence), within the successor theory, cannot in general be established beforehand. For it depends on the details of the successor theory which one ends up finding, and on how much it differs from the original dual models. The successor theory is, however, constrained by the requirement that the originally given dual models must be approximate representations of it, so that the duality is recovered as a special case of, or as an approximation to, the successor theory---and often, it is constrained by additional requirements. Dualities of course often come with indications of the range of parameters for which the successor theory should differ significantly from the given dual models (e.g.~`at strong coupling', for a specific coupling in the theory), and the regime in which it should reproduce the dual models (e.g.~`at weak coupling') or something close to them. 

Dual models often come with specific structure. The theoretical function aims at excising this specific structure, i.e.~finding a theory $T$ that contains only the common core of the dual models. On the other hand, the explicit examples we have reviewed in Section \ref{compare} suggest that the heuristic function {\it makes use of the specific structure}, and reinterprets it as physical structure in the successor theory, $T_{\tn S}$. Therefore, the two functions interpret the models differently.

Thus the set of models, $\{M_i\}_{i\in I}$, and the successor theory, $T_{\tn S}$, are related by a set of dualities and approximation schemes. 
%In the case of 11-dimensional M theory, the approximation is the limit of a small circle, i.e.~an idealisation (of two possible kinds, giving two possible 10-dimensional string theories), and there are duality maps relating the five string theories (S and T dualities). 
Duality and approximation usually relate both the models' and the theory's formalisms (including: the number and nature of the physical degrees of freedom, the dynamics, and the rules for calculating physical quantities) and their interpretation (the model's or theory's reference to worldly items). And so, these are relations between {\it interpreted} theories. 

The schema for duality from De Haro (2016) and De Haro and Butterfield (2017) also illustrates how the two functions of duality can be compatible. We distinguish two theories, $T$ from $T_{\tn S}$, which stand in different relationships to both duality and to the models. The theoretical function aims to construct a theory, $T$, of which the dual models are exact representations. The heuristic function aims to construct the successor theory, $T_{\tn S}$: where the models are {\it not} instantiations, or representations, of the latter theory, but rather approximations to it. In particular, if both the reconstructed theory $T$ and the successor theory $T_{\tn S}$ exist, then one expects that $T$ can be obtained from $T_{\tn S}$ by making the appropriate approximations. Thus, $T$ instantiates duality, while $T_{\tn S}$ does {\it not} instantiate duality: only approximately, through the approximation of its models.

\section*{Acknowledgements}
\addcontentsline{toc}{section}{Acknowledgements}

I thank the members of the research group on Philosophy of Science and Technology of VU University of Amsterdam, and especially Hans Radder and Peter Kirschenmann, for a discussion of the manuscript. I also thank Jeremy Butterfield for comments on the paper and insightful discussions, and especially two referees for their comments. This work was supported by the Tarner scholarship in Philosophy of Science and History of Ideas, held at Trinity College, Cambridge.

\section*{References}
\addcontentsline{toc}{section}{References}

Abraham, R.~and Marsden, J.~E.~(1978). `Foundations of Mechanics'. Addison-Wesley Publishing Company, Inc.\\
\\
Aharony, O., Gubser, S.~S., Maldacena, J.~M., Ooguri, H.~and Oz, Y.~(1999). `Large $N$ field theories, string theory and gravity'. {\it Physics Reports}, 323, 2000, p.~183.\\ doi:10.1016/S0370-1573(99)00083-6  [hep-th/9905111].\\
  %%CITATION = doi:10.1016/S0370-1573(99)00083-6;%%
\\
Banks, T., Fischler, W., Shenker, S.~H.~and Susskind, L.~(1996). `M theory as a matrix model: A Conjecture', {\it Physical Review} D, 55, 1997, p.~5112.   doi:10.1103/PhysRevD.55.5112  [hep-th/9610043].\\
  %%CITATION = doi:10.1103/PhysRevD.55.5112;%%
\\
Belot, G.~(2013). `Symmetry and Equivalence'. Chapter 9 of {\it The Oxford Handbook of Philosophy of Physics}, R.~Batterman (Ed.). Oxford: OUP.\\
\\
Butterfield, J.~(2006). `On symmetry and conserved quantities in classical mechanics'. In: {\it Physical theory and its interpretation}, pp.~43-100. Springer Netherlands.\\
\\
Castellani, E.~and Rickles, D.~(2017). `Introduction to special issue on dualities', {\it Studies in History and Philosophy of Modern Physics}, forthcoming. doi.org/10.1016/j.shpsb.2016.10.004.\\
\\
Caulton, A.~(2015). `The role of symmetry in the interpretation of physical theories'. {\it Studies in History and Philosophy of Science Part B: Studies in History and Philosophy of Modern Physics}, 52, pp.~153-162.\\
\\
Coffey, K.~(2016). `Theoretical Equivalence as Interpretative Equivalence'. {\it The British Journal for the Philosophy of Science}, 65, pp.~821-844.\\
\\
De Haro, S.~(1998). `Planckian scattering and black holes'. {\it Journal of High-Energy Physics}, 9810, 023. doi:10.1088/1126-6708/1998/10/023
  [gr-qc/9806028].\\
  %%CITATION = doi:10.1088/1126-6708/1998/10/023;%%
\\
De Haro, S.~(2015). `Dualities and emergent gravity: Gauge/gravity duality'. {\em Studies in History and Philosophy of Modern Physics}, 59, 2017, pp.~109-125. \\ %doi:~10.1016/j.shpsb.2015.08.004. PhilSci 11666.\\
 %%CITATION = doi:10.1016/j.shpsb.2015.08.004;%%
\\
De Haro, S., Teh, N., Butterfield, J.N.~(2015). `Comparing dualities and gauge symmetries'. {\em Studies in History and Philosophy of Modern Physics}, 59, 2017, pp.~68-80. \\%https://doi.org/10.1016/j.shpsb.2016.03.001\\
%%CITATION = doi:10.1016/j.shpsb.2016.03.001;%%
\\
De Haro, S.~(2016). `Spacetime and Physical Equivalence'. To appear in {\it Space and Time after Quantum Gravity}, Huggett, N.~and W\"uthrich, C.~(Eds.). http://philsci-archive.pitt.edu/13243.\\
%%CITATION = ARXIV:1707.06581;%%
\\
De Haro, S.~(2019). `Towards a Theory of Emergence for the Physical Sciences'. {\it European Journal for Philosophy of Science,} 9, 38.\\
\\
De Haro, S.~(2019a). `Theoretical Equivalence and Duality'. {\it Synthese,} available online: https://doi.org/10.1007/s11229-019-02394-4.\\
\\
De Haro, S.~and Butterfield, J.N.~(2017). `A Schema for Duality, Illustrated by Bosonization'. In: Kouneiher, J.~(Ed.), {\it Foundations of Mathematics and Physics one century after Hilbert}. Springer. \\
%%CITATION = ARXIV:1707.06681;%%
\\
De Haro, S.~and De Regt (2017). `Interpreting Theories without Spacetime'. {\it European Journal for Philosophy of Science}, 8 (3), pp.~631-670.\\
\\
De Regt, H.~W.~(2017). `Understanding Scientific Understanding'. Oxford: OUP.\\
\\
Dieks, D., Dongen, J. van, Haro, S. de~(2014). `Emergence in Holographic Scenarios for Gravity'. 
%PhilSci 11271, arXiv:1501.04278 [hep-th]. 
{\it Studies in History and Philosophy of Modern Physics} 52 (B), 2015, pp.~203-216. doi:~10.1016/j.shpsb.2015.07.007.\\
  %%CITATION = ARXIV:1501.04278;%% 
\\
Dijkgraaf, R.~(1997). `Les Houches lectures on fields, strings and duality'. In {\it Les Houches 1995, Quantum symmetries}, pp.~3-147. [hep-th/9703136].\\
  %%CITATION = HEP-TH/9703136;%%
\\
Fraser, D.~(2017). `Formal and physical equivalence in two cases in contemporary quantum physics'. {\it Studies in History and Philosophy of Modern Physics}, 59, 30-43. \\doi:~10.1016/j.shpsb.2015.07.005.\\
%\\
%Fuchs, J.~and Schweigert, C.~(1997). `Symmetries, Lie Algebras and Representations. A graduate course for physicists'. Cambridge: CUP.\\
\\
Greaves, H., and D.~Wallace (2013). ``Empirical consequences of symmetries''. {\it The British Journal for the Philosophy of Science}, 65 (1), pp.~59-89.\\
\\
Hey, S.~P.~(2016). `Heuristics and Meta-heuristics in Scientific Judgment'. {\it The British Journal for the Philosophy of Science}, 67, pp.~471-495.\\
%\\
%Horowitz, G.~T.~(2005). `Spacetime in string theory'. {\it New Journal of Physics}, 7, p.~201. doi:10.1088/1367-2630/7/1/201  [gr-qc/0410049].\\
  %%CITATION = doi:10.1088/1367-2630/7/1/201;%%
\\
Houkes, W.~and Vermaas, P.E.~(2010). `Technical Functions. On the Use and Design of Artefacts'. Dordrecht Heidelberg London New York: Springer.\\
\\
Huggett, N.~and W\"uthrich, C.~(2013). `Emergent spacetime and empirical (in)coherence'. {\it Studies in History and Philosophy of Modern Physics}, 44, pp.~276-285.\\
\\
Huggett, N. (2017), `Target space $\neq$ space'. {\em Studies in History and Philosophy of Modern Physics}, 59, 81-88. doi:10.1016/j.shpsb.2015.08.007.\\
\\
Kempf, A.~(1994). `Uncertainty relation in quantum mechanics with quantum group symmetry'.
  {\it Journal of Mathematical Physics}, 35, p.~4483
  doi:10.1063/1.530798
  [hep-th/9311147].\\
  %%CITATION = doi:10.1063/1.530798;%%
\\
Laudan, L.~(1984). `Science and Values. The aims of science and their role in scientific debate'. Berkeley Los Angeles London: University of California Press.\\
\\
Maggiore, M.~(1993). `A Generalized uncertainty principle in quantum gravity'. {\it Physics Letters} B 304, p.~65 doi:10.1016/0370-2693(93)91401-8
  [hep-th/9301067].\\
%%CITATION = doi:10.1016/0370-2693(93)91401-8;%%
Matsubara, K.~(2013). `Realism, underdetermination and string theory dualities'. {\it Synthese}, 190 (3), pp.~471–489.\\
\\
Nagel, E.~(1961). {\it The Structure of Science: Problems in the Logic of Scientific Explanation}. New York: Harcourt.\\
\\
Polchinski, J.~(1998). `String theory. Volume 1: An introduction to the bosonic string'. Cambridge: CUP.\\
\\
Psillos, S.~(1999). {\it Scientific Realism. How Science Tracks Truth.}  London and New York: Routledge.\\
\\
Radder, H.~(1991). `Heuristics and the Generalized Correspondence Principle'. {\it The British Journal for the Philosophy of Science}, 42, 195-226.\\
\\
Radder, H.~(2012). `Postscript 2012', in: {\it The Material Realization of Science}, Revised Edition, Springer Dordrecht Heidelberg New York London.\\
\\
Read, J.~(2016). `The Interpretation of String-Theoretic Dualities. {\it Foundations of Physics}, 46, pp.~209-235.\\
\\
Rickles, R.~(2011). `A Philosopher looks at String Dualities'. {\it Studies in History and Philosophy of Modern Physics}, 42 pp.~54-67.\\
\\
Rickles, D.~(2013). `AdS/CFT duality and the emergence of spacetime'. {\it Studies in History and Philosophy of Modern Physics}, 44, pp.~312-320.\\
\\
Rickles, D.~(2017). `Dual theories: `same but different' or different but same'?' {\em Studies in History and Philosophy of Modern Physics}, 59, 62-67. doi:~10.1016/j.shpsb.2015.09.005.\\
\\
Ruetsche, L.~(2011). {\it Interpreting Quantum Theories}. Oxford University Press.\\
%\\
%Sitartz, A.~(1995). `Noncommutative geometry and gauge theory on discrete groups'.  {\it Journal of Geometry and Physics}, 15, p.~123.   doi:10.1016/0393-0440(94)00009-S
  %[hep-th/9210098].\\
  %%CITATION = doi:10.1016/0393-0440(94)00009-S;%%
\\
Snyder, H.~S.~(1947). `Quantized Space-Time'. {\it Physical Review}, 71 (1), pp.~38-41.\\
\\
Teh, N.~J.~(2016). ``Galileo’s gauge: Understanding the empirical significance of gauge symmetry''. {\it Philosophy of Science}, 83 (1), pp.~93-118.\\
\\
Toulmin, S.~(1961). `Foresight and Understanding. An enquiry into the aims of Science'. Indiana University Press.\\
\\
Whewell, W.~(1876). Letter to Prof.~J.~D.~Forbes (1960), in {\it William Whewell, D.~D., Master of Trinity College, Cambridge: an account of his literary and scientific correspondence}, vol.~2, Todhunter, I. London: Macmillan 1876.\\
\\
Witten, E.~(1995). `String Theory Dynamics in Various Dimensions'. {\it Nuclear Physics B}, 443 (1-2), pp.~85-126.\\

\end{document}